\documentclass[acmtecs]{acmsmall}
\usepackage[ruled]{algorithm2e}
\usepackage{verbatim}
\usepackage{lscape}

\usepackage{subfigure}
\usepackage{amssymb}
\usepackage{amsmath}
\begin{document}

% Page heads
%\markboth{G. Zhou et al.}{A Multifrequency MAC Specially Designed for WSN Applications}

% Title portion
\title{Leveraging Subjective Human Annotation for Clustering Historic Newspaper Articles}
\author{Haimonti Dutta
\affil{The Center for Computational Learning Systems}
William Chan
\affil{Department of Computer Science}
Deepak Shankargouda
\affil{Department of Computer Science}
Manoj Pooleery
\affil{The Center for Computational Learning Systems}
Axinia Radeva
\affil{The Center for Computational Learning Systems}
Kyle Rego
\affil{Department of Computer Science}
Boyi Xie
\affil{The Center for Computational Learning Systems}
Rebecca J. Passonneau
\affil{The Center for Computational Learning Systems}
Austin Lee
\affil{The Center for Computational Learning Systems}
Barbara Taranto
\affil{New York Public Library}}

\begin{abstract}

%\begin{comment}

The New York Public Library is participating in the \textit{Chronicling America} initiative to develop an online searchable database of historically significant newspaper articles. Microfilm copies of the newspapers are scanned and high resolution Optical Character Recognition (OCR) software is run on them. The text from the OCR provides a wealth of data and opinion for researchers and historians. However, categorization of articles provided by the OCR engine is rudimentary and a large number of the articles are labeled ``\textit{editorial}" without further grouping. 
Manually sorting articles into fine-grained categories is time consuming if not impossible given the size of the corpus. This paper studies techniques for automatic categorization of newspaper articles so as to enhance search and retrieval on the archive. We explore unsupervised (e.g. KMeans) and semi-supervised (e.g. constrained clustering) learning algorithms to develop article categorization schemes geared towards the needs of end-users. A pilot study was designed to understand whether there was unanimous agreement amongst patrons regarding how articles can be categorized. It was found that the task was very subjective and consequently automated algorithms that could deal with subjective labels were used. While the small scale pilot study was extremely helpful in designing machine learning algorithms, a much larger system needs to be developed to collect annotations from users of the archive. The ``BODHI" system currently being developed is a step in that direction, allowing users to correct wrongly scanned OCR and providing keywords and tags for newspaper articles used frequently. On successful implementation of the beta version of this system, we hope that it can be integrated with existing software being developed for the Chronicling America project.

\end{abstract}




\acmformat{Dutta, H., Passonneau, R. J., Chan, W., Xie, B., Pooleery, M., Radeva, A., Shankargouda, D., Rego, K., Lee, A., Taranto, B. 2012. Leveraging Subjective Human Annotation for Clustering Historic Newspaper Articles.}

\begin{bottomstuff}
%This work is supported by the National Endowment for Humanities, under
%grant NEH HD 51153-10. 

Author's addresses: H. Dutta {and} R. J. Passonneau {and} B. Xie {and} M. Pooleery {and} A. Radeva, The Center for Computational Learning Systems, Columbia University,
New York, NY 10115; W. Chan  {and} D. Shankargouda {and} K. Rego,
Department of Computer Science, Columbia University; A. Lee, SyncSort Inc. This work was done when A. Lee was affiliated to The Center for Computational Learning Systems, Columbia University; B. Taranto, New York Public Library, NYPL.
\end{bottomstuff}

\maketitle

\section{Introduction}

\noindent \textit{Chronicling America} (\texttt{http://chroniclingamerica.loc.gov/}) is a website that provides access to over 3 million historic newspapers scanned by the National Digital Newspaper Program (NDNP). It is an initiative of the National Endowment for Humanities (NEH) and the Library of Congress (LC) and its goal is to develop an online, searchable database of historically significant newspapers between 1836 and 1922. State libraries, historical societies and universities have been funded by the NEH to generate scanned newspaper pages representing the state's regional history, geographic coverage, and events which will be archived by the LC. The New York Public Library (NYPL) is part of this initiative and has scanned 200,000 newspaper pages published between 1860 and 1920 from microfilm. 
%The opening of the Brooklyn Bridge in 1883, the construction of an immigration station at Ellis Island in 1890 and the historic opening of the Metropolitan Opera House on Broadway and 39th Street are examples of some interesting news items of the time.

%Interestingly, once digitized, the archive is The scanned newspaper holdings of the NYPL offers a wealth of data for researchers, historians, genealogists and other patrons of the library. The team at NYPL is therefore interested in enabling easy search and retrieval techniques on its archive. 

%The goal of our research project is to enable users of this historical archive including scholars (genealogists, geologists, marine biologists investigating oil spills in the New York area) and other library patrons to efficiently search for articles of interest to them. 

While the images scanned by the NYPL are sent to the Library of Congress for hosting on the Chronicling America website, duplicate copies of the archive are also available at NYPL for use by its patrons. These digitized pages offer a wealth of data for researchers, historians, genealogists and other patrons of the library. For example, the opening of the Brooklyn Bridge in 1883, the construction of an immigration station at Ellis Island in 1890, the historic opening of the Metropolitan Opera House on Broadway and 39th Street are some interesting local news items of the time. Other topics widely covered by the American press include presidential administrations (Cleveland (1885 -- 1894), Garfield (1880 - 1883), McKinley (1897 - 1901), Theodore Roosevelt (1901 - 1912)), natural calamities (Galveston flood of 1900, San Francisco earthquake 1906), the sinking of the Titanic, events pertaining to the first world war and news from the world of medicine (patented medicines, spread of epidemics, new discoveries). To effectively use this archive, developing sophisticated search and retrieval mechanisms is crucial.

\noindent In order to make a newspaper available for searching on the Internet, the following processes must take place: 
\begin{enumerate}
\item The microfilm copy or original paper is scanned.
\item Master and Web image files are generated. 
\item Metadata is assigned for each page to improve the search capability of the newspaper.
\item Optical Character Recognition (OCR) software is run over high resolution images to create searchable full text, and 
\item OCR text, images, and metadata are imported into a digital library software program. 
\end{enumerate}

\noindent The newspaper archives can currently be searched using the OpenSearch protocol\footnote{\texttt{http://www.opensearch.org/Home}}. Unfortunately, these search facilities are rudimentary and irrelevant documents are often more highly ranked than relevant ones. For instance consider a search for a natural calamity like the April 18th, 1906 San Francisco earthquake which killed approximately 2000 people and measured 7.8 on the Richter scale; if the keywords ``earthquake San Francisco" are entered as the search criteria in the digitized newspaper archive along with a date range 01/01/1906 to 12/31/1906, the first document returned is Page 7, April 19th of the 1906 Los Angeles Herald with an article ``Fifty people killed at San Jose" (the word ``San" is tagged); the second document returned is the June 3rd, 1906 issue of ``The San Francisco Sunday Call" with a full-page illustration of a drama by Frederick Irons Bamford and the third document is page 13, June 3rd, 1906 issue of ``The Sunday Call" with a cartoon of ``Major ozone$'$s fresh air crusade". The retrieval technique missed finding Page 1, April 19th, 1906 of the newspaper ``The Sun" published from New York, which had a headline article ``Earthquake lays Frisco in ruins". 

\begin{table}[h]
\tbl{The Original Text in the article versus the scanned OCR.
\label{garbageOCR}}
{
%\begin{center}
\begin{tabular}{|c || c|} \hline
ARMY BILL PROVISIONS. & JLRMY BILL PROVISIONS \\ \hline 
INCREASE AND REORGANIZATION & INCREASE AND REORCAMZATICN \\ 
RANKS MAY REACH ONE HUNDRED THOU- & HANKS MAT REACH ONE HUNDRED THOU \\ 
SAND IN EMERGENCIES-FILIPONOS AND & SAND IN EMERGENCIES Ai FILIPINOS AND  \\ 
PORTO RICANS TO BE ENLISTED. & PORTO RICANS TO BE ENLISTED. \\  \hline 
[BY TELEGRAPH TO THE TRIBUNE] & IBT TCLEGSAPB TO TCS TRIBUNE] \\ 
Washington, Nov. 30-The approved Army In- & ``Washington, Nov. Ai-he approved Army in- \\ 
crease and Reorganization bill agreed upon by & crease and Reorganization bill agreed u(>on by \\ 
Secretary Root and General Corbin on the one & Secretary Root and General Corbin on the one \\ 
hand and the Senate and House Military com- & hand and the Senate and House Military com- \\
mittees on the other is an elaborate measure of & rr.ittees on the other is an elaborate measure of \\ 
thirty-nine sections, providing for a detailed in & thirty-nine sections, providing for a detailed in \\
stead of a permanent staff; for an artillery & stead of a permanent staff; for an artillery \\ 
corps, coast and field, instead of regiments; for & corps, coast and field, instead of regiments; for  \\
an Army of 60,000, including officers to be in- & en Army of Go.<"-"including of3eers to be in \\
creased in ranks to 100,000 in emergencies, with & creased in ranks to 100.000 in emergencies, with \\
Congressional approval; & Congressional approval; \\ \hline
\end{tabular}
%\end{center}
}
%\label{default}
\end{table}%

On investigating the reasons why irrelevant documents are ranked higher, it was found that the newspapers are scanned on a page-by-page basis and article level segmentation is poor or non-existent. The OCR scanning process is far from perfect and the documents generated from it contains a large amount of garbled text. Table~\ref{garbageOCR} (Left) shows the text in an article and (Right) the garbled text generated by the scanning process. 
%(TALK ABOUT THE SCAN CONFIDENCE OF WORDS)
In addition, categorization of article level data using the OCR software was not very successful -- most of the articles are labeled ``editorial" and there is no fine grained classification into categories such as crime, politics, medicine, etc. For example, an attempt to categorize articles in the edition of \textit{The Sun} newspaper published on November 4th, 1894 resulted in 338 articles classified as editorial, 32 unclassified\footnote{These were later identified as banners of the newspaper.}, 10 sports, 23 advertising, 5 commercial, 3 birth-related announcements, and 2 reviews.
% (POSSIBLY PROVIDE A HISTOGRAM OF ARTICLES IN OUR ARCHIVE)

Even though the New York Public Library has put in substantial effort into improving the quality of the images and text obtained from OCR by testing each word scanned against english dictionaries, manually re-typing newspaper headlines and applying categories to articles  -- the digital outputs from the OCR software are not good enough to ensure adequate quality of text retrieval or to meet user expectations. Consequently, we conjectured that a crowdsourcing project involving patrons of the library who use the archive frequently for research and learning could assist the task of categorizing articles of this huge online repository.  

%Motivated by a similar project at the National Library of Australia(\texttt{http://trove.nla.gov.au}), the team at NYPL decided to ``crowdsource" different sub-tasks in the project such as enhancing the data quality from garbled OCR and providing relevant tags and keywords for articles. Patrons of the library are happy to become involved with a project that would benefit the ``common good" and also help with enhancing a valuable and rich resource -- the online historical newspaper archive. 

This paper describes our experiences when setting up a pilot study on a randomly chosen sample of newspaper articles from the archive. Annotators were asked to browse the articles and come up with broad categories into which they thought the articles could be grouped; they were not given a pre-defined list of categories to choose from. An attempt was made to leverage the information provided by them (such as keywords and tags, labels) and study whether popular unsupervised\footnote{Since well-defined labels are not easily available from the archive, unsupervised machine learning algorithms were preferred to the well known supervised counterparts.} machine learning and text mining algorithms benefited from such prior knowledge. In addition, annotators are evaluated based on the degree to which the additional knowledge they provide helps the learning task. 
%Finally, we present an architecture for the OCR corrector and tag collector that are currently being integrated into the library environment for a large scale study. 

\begin{comment}
describes problems faced by annotators in assigning categories to documents when there were no pre-defined when applying sophisticated text mining and machine learning algorithms on the archive including presence of multiple annotators providing class labels or categories of articles, no ground truth observed, users able to suggest different number of categories if this was not pre-defined,  how to apply unsupervised learning algorithms (such as clustering in this setting). Particular emphasis is given to evaluate the quality of annotations obtained and understand whether the existence of multiple labels and prior domain knowledge was able to improve the accuracy of machine learning algorithms. 
\end{comment}

This paper is organized as follows: Section~\ref{related} presents related work, Section~\ref{data} describes the characteristics of the data, Section~\ref{pilot} provides details of the pilot study and interprets the results, Section~\ref{unsupervised} describes algorithms for incorporating domain knowledge into learning tasks and empirical analysis performed on this data, Section~\ref{tagworks} describes the implementation of a system for correcting OCR text and collecting tags and Section~\ref{conc} concludes the paper.

\section{Related Prior Work}
\label{related}
In recent years, the humanities have seen a transformation of scholarly information from physical media to digital form, resulting in the formation of large digital libraries (such as ARTstor\footnote{\texttt{www.artstor.org}}, Data-PASS\footnote{\texttt{http://www.icpsr.umich.edu/DATAPASS/}}, Biodiversity Heritage Library\footnote{\texttt{biodiversitylibrary.org}}, English Broadside Ballad Archive\footnote{\texttt{ebba.english.ucsb.edu}}, Great War Primary Documents Archive\footnote{\texttt{www.gwpda.org}}, National Archives of London\footnote{\texttt{nationalarchives.gov.uk}} ). 
%In April 2007, the US National Science Foundation (NSF) and the British Joint Information Systems Committee (JISC) held an invitational workshop on data-driven scholarship that aimed to identify opportunities and strategies for managing information created by researchers and scholars. 
%A new form of research, called \textit{cyberscholarship} (\cite{Arms_07}) has emerged when high performance computing meets online repositories, a. 
Vast quantities of datasets, manuscripts, reports and newspapers that might never have made their way into a traditional library are being made accessible to the general public using high-performance computing, networking and storage. 
%The basic idea driving this research is that a scholar reads only a few hundreds of documents, but a supercomputer or a network of machines can analyze millions. 

Several digital humanities projects that have used machine learning and natural language processing techniques to learn from historic newspaper archives are relevant to this work -- the libraries of Richmond and Tufts have examined the \emph{Richmond Times Dispatch} during the civil war years for more than two decades and their work focuses on automatic identification and analysis of full OCR text in newspapers to provide advanced searching, browsing and visualization \cite{Crane_06}. The focus of this work was on named entity extraction and ten categories prominent in these newspapers were studied including ship names, railroads, streets and organizations. In an earlier project at the universities, the Perseus project (\cite{Smith_02}, \cite{Smith_02a}, \cite{Smith_01a}), a general system to extract dates and names from text was developed. At the Hull Digital Library\footnote{\texttt{http://www2.hull.ac.uk/lli/libraries.aspx}}, information capture and semantic indexing of the newspaper archive is done using document analysis and classification \cite{Esposito_97}. In Collection OCR, Sankar et. al. \cite{SankarK_10} use an approximate fast nearest neighbor algorithm based on hierarchical K-Means (HKM) to clean OCR text. In general, the focus of almost all of the prior work in digital humanities has been on language modeling and has not focused on categorization of articles taking into consideration subjectivity of human annotation. Finally, it must be noted that a preliminary version of this work using unsupervised learning algorithms was published in \cite{Dutta_11a} and the problem of topic evolution in the historic newspaper archive over time was explored by Lee et. al. \cite{Dutta_11b}.

\noindent Recruiting web users to tag and annotate text and images in large archives, especially when there are too many documents and objects for a single authority to label has become common practice. Several projects use the collective effort of a large number of people to label text, images and annotate maps; with the advent of crowdsourcing services (such as Amazon's Mechanical Turk\footnote{\texttt{https://www.mturk.com/mturk/welcome}}, reCAPTCHA (\cite{vonAhn_08}, \cite{Faymonville_2009}), the LISTEN (\cite{Turnbull_07}), ESP (\cite{vonAhn_04}) and Games with a Purpose (\cite{Ahn_08}) paying people small sums of money to do ``Human Intelligence Tasks" (HITs) is also becoming the norm. Such tasks include anything from labeling images, to listening to short pieces of audio, researching topics on the internet and scrubbing database records. A number of recent papers have evaluated the effectiveness of using Mechanical Turk to create annotated data for text and natural language processing applications (see for e.g. \cite{Snow_08}, papers accepted at the ``Creating Speech and Language Data With Amazon's Mechanical Turk" Workshop\footnote{http://sites.google.com/site/amtworkshop2010/home} co-located with HLT-NAACL 2010). While it is relatively inexpensive to obtain a large number of labels from annotators, the question of how to incorporate them into machine learning algorithms with theoretical guarantees on performance remains an open research problem. 
%\end{comment}
%WRITE A FEW SENTENCES HERE REGARDING THE ABOVE PARAGRAPH. 

In the context of supervised learning tasks, Smyth et al. were the first in the machine learning community to propose a solution to the problem of noisy labels in the context of labeling volcanoes in satellite images of Venus (\cite{Smyth_94}, \cite{Burl_94}, \cite{Smyth_96}). They first estimate the ground truth and then use probabilistic reasoning to learn the classifier. Raykar and his colleagues \cite{Raykar_09} describe a probabilistic approach when multiple annotators provide possibly noisy labels but there is no absolute gold standard. Their algorithm iteratively establishes a particular gold standard, measures the performance of annotators given that gold standard and then refines it based on performance metrics. Key assumptions made by them include: a) performance of each annotator does not depend on the feature vector and b) conditioned on the truth the experts are independent. 

\cite{Sheng_08} analyzed when it is worthwhile to acquire new labels for some training examples. They show that repeated labeling can improve label quality, but not always; when labels are noisy, repeated labeling can be preferable to single labeling even in the traditional setting where labels are not particularly cheap. An empirical study to examine the effect of noisy annotations on the performance of sentiment classification models was performed by \cite{Hsueh_09}. More theoretical work on when it is useful to deal with multiple experts can be found in \cite{Lugosi_92}, \cite{Dekel_09}, \cite{Dekel_09a}. 

Another class of algorithms that needs discussion are \textit{semi-supervised} clustering algorithms. These algorithms fall in between totally unsupervised learning and totally supervised learning. The primary goal of such algorithms is to ``steer" the clustering process with user feedback; also the clusters obtained by guidance from humans enable the users to play with the data and understand it intuitively. Semi-supervised clustering algorithms can be broadly classified into two main genres: semi-supervised clustering of (a) labels and (b) constraints (\cite{Basu_05}). 

Using labeled data, iterative feedback from users has been incorporated by \cite{Cohn_03} and methods for using conditional distributions in auxiliary space are reported in \cite{Sinkkonen_02}. Seeding mechanisms for semi-supervised clustering have been studied by \cite{Basu_02} and \cite{Wagstaff_01}. For iterative clustering algorithms (such as K-Means), a common technique is to seed at random by arbitrarily creating $K$ partitions and choosing the mean of each partition as seeds. \cite{Forgy_65} proposed a variant that chooses $K$ instances at random as seeds, then assigns the remaining instances to the cluster represented by the nearest seed. \cite{MacQueen_67} recalculates the centroids after the assignment of instances to the cluster represented by the nearest seed. \cite{Kaufman_90} propose an elaborate mechanism of seed selection: the first seed is the instance that is most central in the data; the rest of the representatives are selected by choosing instances that promise to be closer to more of the remaining instances. Other interesting seeding mechanisms include the Buckshot method of doing hierarchical clustering on a sample of data to get initial set of cluster centers (\cite{Cutting_1992}), selecting the $k$ densest intervals along each co-ordinate to get the $k$ cluster centers (\cite{Bradley_97}) and refining the initial seeds by taking into account the modes of the underlying distribution (\cite{Bradley_98}). In K-Means++ (\cite{Arthur_07}), the random starting points are chosen with specific probabilities. By augmenting K-Means using this simple, randomized seeding technique, K-Means++ is $\theta$ (log $K$) competitive with the optimal clustering. 

In semi-supervised clustering with constraints, the focus is on either: (a) similarity adapting or (b) search-based methods. In similarity-adapting methods, an existing clustering algorithm using some similarity measure is employed, but the similarity measure is adapted to suit the problem (such as the Jensen-Shannon divergence trained with gradient descent \cite{Cohn_03}, the Euclidean distance modified by a shortest-path algorithm \cite{Klein_2002}, Mahalanobis distances adjusted by convex optimization \cite{Bilenko_2003}, \cite{Xing_02}. In search-based methods, the clustering algorithm itself is modified so that user-provided constraints can be used to bias the search for an appropriate clustering. This can be done in several ways, such as by performing a transitive closure of the constraints and using them to initialize clusters \cite{Basu_02}, by including in the cost function a penalty for lack of compliance with the specified constraints \cite{Demiriz_99}, or by requiring constraints to be satisfied during cluster assignment in the clustering process \cite{Wagstaff_01}. 

Pairwise constrained semi-supervised clustering has also been studied by \cite{Bansal_04}, \cite{Blum_04}, \cite{Kulis_09}, \cite{Lange_05}, \cite{Ge_07}, \cite{Davidson_06a}) and \cite{Charikar_03} and a related approach based on Gaussian random field model is presented by \cite{Zhu_03}. A probabilistic model for semi-supervised clustering based on Hidden Markov Random Fields was studied by \cite{Basu_04} where the goal is to perform partitional semi-supervised clustering of data by minimizing an objective function derived from the posterior energy of the HMRF model. 

%In penalized probabilistic clustering (\cite{Lu_07}) clustering preferences are expressed in a prior distribution over assignments of data points to clusters; this prior penalizes cluster assignments according to the degree with which they violate the preferences and the model parameters are fit with the expectation-maximization (EM) algorithm. It is worth mentioning that the proposed research 
%is different from \emph{active} semi-supervised learning (\cite{Basu_04}, \cite{Grira_05}, \cite{Grira_08}, \cite{Xue_09}) where the assumption is that not all of the labeled data is available initially, but the algorithm can pose questions regarding whether must-link or cannot-link constraints exists between two data points from a noiseless oracle.

% constraint-based or distance based clustering. COP-KMeans (\cite{Wagstaff_01}) has a heuristically motivated objective function and pairwise constrained clustering is studied by Bansal et. al, Blum et. al. and Charikar et. al.; distance based methods are studied by Xing et. al., Redundant Component Analysis, and must-link constraints to learn a Mahalanobis distance using convex optimization. 

\cite{Cohn_03} note that ``semi-supervised clustering assumes that human user has in their mind criteria that enable them to evaluate the quality of clustering". It does not assume that the user is conscious of what they think defines a good clustering but that, as with art, they will ``know it when they see it". While this is very interesting, an important observation is that the \textit{subjectivity of labels} assigned by humans is not discussed in the context of evaluation of semi-supervised clustering; furthermore, several open questions exist including incomplete seeding techniques (what if some cluster labels are not observed in the labeled set available) and how to deal with noise in the data.

\section{The Data}
\label{data}
%\subsection{Description}

\begin{figure}[t]
\begin{center}
\includegraphics[height=0.45\textheight]{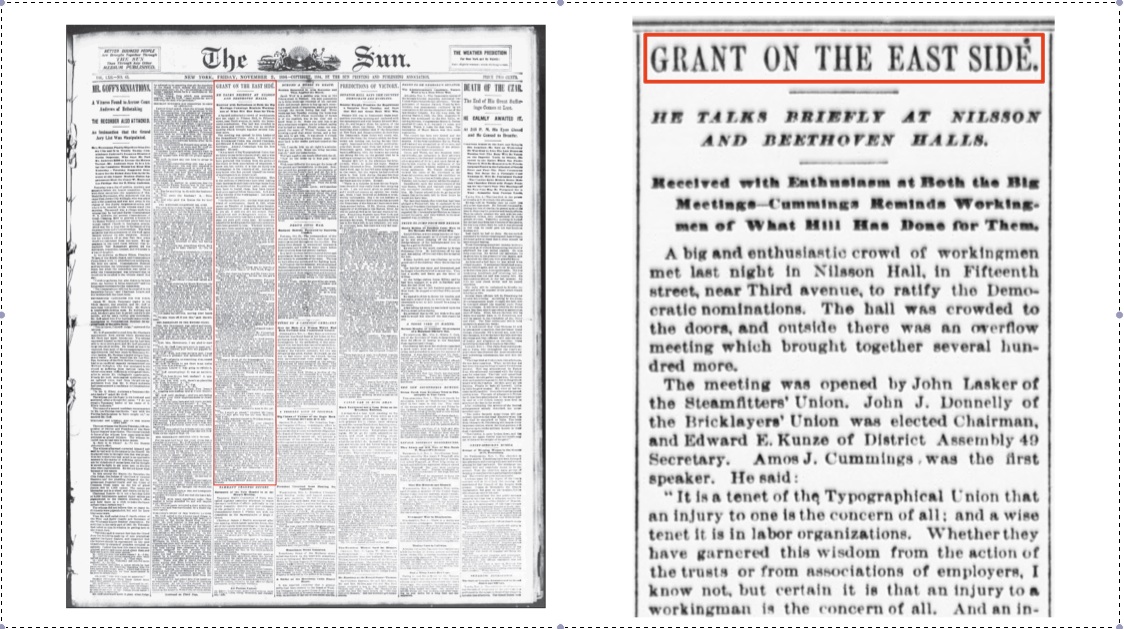}
\caption{(Left) A newspaper page from the NYPL archive. The red-border shows an article from the newspaper, zoomed in on the right hand figure.}
\label{news}
\end{center}
\end{figure}

\begin{figure}[h]
\begin{center}
\includegraphics[width=\textwidth, height=1.7in]{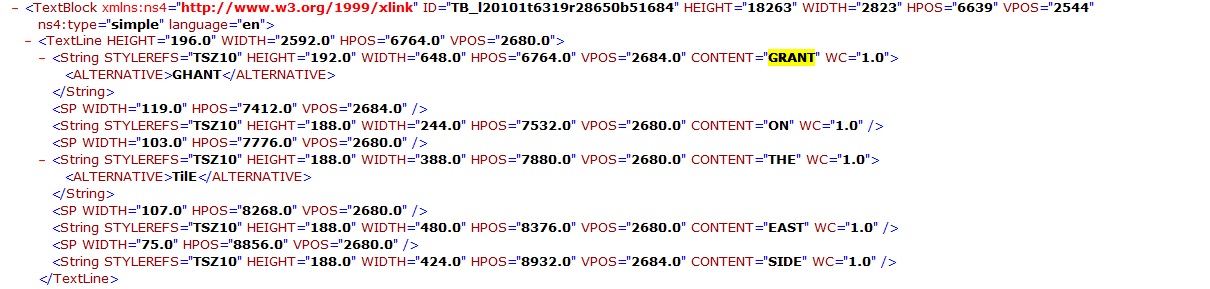}
\caption{Page Level XML file showing the article with headline ``GRANT ON THE EAST SIDE".}
\label{pageLevel}
\end{center}
\end{figure}

\begin{figure}[h]
\begin{center}
\includegraphics[width=0.75\textwidth]{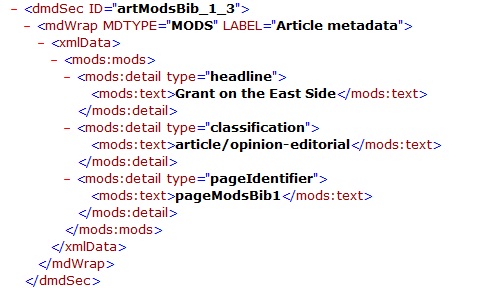}
\caption{A segment of the issue-level XML file illustrating the OCR Classification (as ``article/editorial") for the article and its headline.}
\label{issueLevel}
\end{center}
\end{figure}

Figure~\ref{news} shows a scanned newspaper page (\textit{The Sun}, November 2, 1894) from the NYPL archive and an article from this paper. The historical newspaper archive contains two types of XML files: (1) \textbf{Page-Level XMLs:} For each page of a newspaper, there is an XML file that contains metadata about the page and the text in it. Each word scanned is stored as a string in the page level XML (See Figure~\ref{pageLevel}) along with possible alternative suggestions for the word\footnote{We found that its primary selections are usually better than their alternatives.}. The text is extracted from the page level XML using Xpath queries and stored in a PostGreSQL database. 
%[TALK A BIT ABOUT VERSIONING MECHANISM ONCE THE TEXT GETS CORRECTED; ALSO DISCUSS ARTICLE SEGMENTATION OVER COLUMNS IN SOME DETAIL; ACKNOWLEDGE THAT ARTICLES SPANNING MULTIPLE PAGES ARE NOT DEALT WITH]
(2) \textbf{Issue-Level XMLs:} The issue-level XMLs (illustrated in Table~\ref{issueLevel}) provide the following information about articles: (a) \emph{Headlines cleaned by humans} which are of much higher quality than the text produced by the OCR software. (b) \emph{Article segmentation information:} Each newspaper article is represented as a collection of one or more text blocks whose pixel coordinates are available. This helps to determine where one article ends and the next one begins and is particularly useful when an article spans more than one page. (c) \emph{High-level categorization} of the articles produced by the OCR software.
We have access to only a subset of the NYPL archive\footnote{These have been substantially cleaned by humans} -- issues of \textit{The Sun} newspaper from November 1, 1894 to December 31, 1894. (d) In addition, issue level XMLs also store 
the date of publication, volume number and issue number and provide pointers to the location and names of the page-level XML files. 

%For experiments used in this paper, we used only one randomly chosen newspaper (November 2nd, 1894 issue of \textit{The Sun}). 

Table~\ref{catTab} shows all the categories found by the OCR software for ``The Sun" newspaper published between November 2nd, 1894 -- December 31st, 1894. Articles in the ``editorial/opinion" and ``sports" categories contain statistically significant amounts of text while reviews, illustrations, birth/death/wedding announcements are not included in our study.

%\begin{table}
%\tbl{A segment of the issue-level XML file illustrating the OCR Classification (as ``article/editorial") for the article and its headline.\label{issueLevel}}
%{
%\begin{tabular}{l}
%$<$dmdSec ID=``artModsBib$\_1\_3$" \\
%  $<$mdWrap MDTYPE=``MODS" LABEL=`` \\
%    Article metadata"$>$ \\
%    $<$xmlData$>$ \\
%    $<$mods:mods$>$\\
%$<$mods:detail type=``\textbf{headline}"$>$\\
%$<$mods:text$>$\textbf{Grant on the East Side}$<$/mods:text$>$\\
%$<$/mods:detail$>$\\
%$<$mods:detail type=``\textbf{classification}"$>$\\
%$<$mods:text$>$\textbf{article$/$opinion-editorial}$<$/mods:text$>$\\
%$</$mods:detail$>$\\
%$<$mods:detail type=``pageIdentifier"$>$\\
%$<$mods:text$>$pageModsBib1$</$mods:text$>$\\
%$</$mods:detail$>$\\
%$</$mods:mods$>$\\
%$</$xmlData$>$\\
%$</$mdWrap$>$\\
%$</$dmdSec$>$\\
%\end{tabular}
%}
%%\caption{A segment of the issue-level XML file illustrating the OCR Classification (as ``article/editorial") for the article and its headline.}
%%\label{issueLevel}
%\end{table}

\noindent \textbf{Pre-processing:} We preprocess the documents to reduce dimensionality and have clean data to learn from. For each article, a bag-of-words representation and tf-idf weights are obtained. Stop words such as ``the'', ``and'', etc. are removed from the set of words. Letters of length three or less and words that contain digits or repeated characters (e.g. ``paaa'' and ``ornnn'') are also removed. After applying the above noise reduction techniques, the dimensionality of the feature space is 3210. 
%An important attribute that we included in our study was the number of lines in each article. 
%Figure~\ref{LinCount} illustrates the distribution of line number counts for articles in the repository which seems to suggest that they follow a Zipfian distribution. 

\begin{figure}[t]
%\begin{minipage}[b]{0.5\linewidth}
\centering
%\begin{table}
\begin{tabular}{l|r}
\hline
Category & Article Counts \\ \hline
Editorial/Opinion & 11,441 \\ 
Sports & 764\\
Advertising & 683 \\ 
Commercial/Legal/Public notices & 361 \\ 
Birth/Death/Wedding & 158 \\ 
Reviews & 45 \\
Illustrations & 2 \\
Unclassified & 785 \\ 
Total & 14,239 \\ \hline
\end{tabular}
\caption{Top-level categories of articles from OCR  software for \textit{The Sun} newspaper between November 2nd, 1894 and December 31st, 1894.}   
\label{catTab}
%\end{table}
%\end{minipage}
%\hspace{0.4cm}
%\begin{minipage}[b]{0.5\linewidth}
%\centering
%\includegraphics[totalheight=2.1in]{hist01.jpg}
%\caption{Distribution of article line number counts}
%\label{LinCount}
%\end{minipage}
\end{figure}

%\begin{figure}[htbp]
%\begin{center}
%\includegraphics[width=0.5\textwidth,height=0.3\textheight]{cat.jpg}
%\caption{Histogram of sub-categories found by human annotators in a random sample of 25 articles labeled ``editorial" by the OCR software.}
%\label{catCount}
%\end{center}
%\end{figure}

\section{Case Study Involving Human Annotators}
\label{pilot} 
We conducted a pilot study to test whether the category labeled ``article/editorial" by the OCR software could be further broken down to more meaningful sub-categories. Six annotators were recruited to determine
the number of natural categories found in a random sample of twenty-five articles. The
articles (all labeled �article/editorial� by the
OCR software) were selected from the November 2nd, 1894 issue
of \textit{The Sun} newspaper. All the annotators were given the same set of articles to work with. They were asked to skim the articles first and group them into obvious and intuitive categories and focusing on the ``bigger picture". The defined categories had be described in 5 - 10 words and preferably had to include words from the articles. Finally, they were interviewed with the following set of questions:
\begin{enumerate}
\item What was the strategy you used for coming up with the categories?
\item Were there any documents that you found difficult to assign to categories?
\item Did you find any part of the study particularly difficult or ambiguous? If so, describe the problem you faced.
\item How long did it take you to complete the study?
\item If you had the opportunity to change anything with this study, what would it be?
\end{enumerate} 
While there are many other interesting research questions that can be investigated with human annotated data, the focus was on determining a meaningful number of sub-categories for the ``article/editorial" category; thus reaction times, self-consistency among annotators were not emphasized. 

\begin{table}
\begin{center}
\tbl{Sub-Categories found by humans in the random sample of 25 articles labeled ``article/editorial" by the OCR software.\label{catCount}}
{
\begin{tabular}{l|r}
\hline
 ID & Number of sub-categories \\\hline
Annotator 1 & 8 \\ 
Annotator 2	& 14 \\
Annotator 3	& 13 \\
Annotator 4	& 9 \\
Annotator 5	& 10 \\
Annotator 6	& 9 \\\hline
\end{tabular}
}
%\caption{Sub-Categories found by humans in the random sample of 25 articles labeled ``article/editorial" by the OCR software.}
%\label{catCount}
\end{center}
\end{table}

\subsection{Interpreting Results from the Pilot Study}
\label{interpret}
 Table~\ref{catCount} shows the number of categories found by the annotators. The November 2nd, 1894 newspaper was published immediately after general elections; thus a lot of articles in this issue had to do with politics and elections. This is also reflected in the random sample used for the categorization task -- annotators unanimously agreed that seven of the twenty five articles used for the study belong to the category ``politics/elections/governmental appointments". Three of the annotators found hierarchies among this category such as ``politics/ballot, politics/election, politics/nomination, politics/war, politics/social, politics/entertainment, politics/gossip". This accounted for the increased number of total categories they listed. Since the instructions explicitly mentioned focusing on the ``bigger picture" and not drilling down to very fine-grained categories, these were merged together to form the category ``politics". Annotators also agreed unanimously on one article belonging to the category ``medicine, public-health, and safety". This article presented a report on a new diphtheria remedy and announced the arrival of fresh serum from Germany which was tried on two cases in Philadelphia. They merged together articles that contained arts, biographies, book reviews and the like into one category called ``arts/human interest". Creating a homogeneous category for these articles was not easy due to the wide variety of articles. Articles pertaining to ``death" and ``marriage announcements" were binned into separate categories. There was no agreement among annotators on eleven articles -- for example, an article with a headline ``President Cleveland goes hunting for squirrels" was labeled as belonging to the following categories: human interest/politics/sports/entertainment/social. All of these eleven articles had a much higher level of ambiguity and there was no agreement among annotators. Since we did not have categories pre-defined for the annotation task and chose rather to let annotators come up with appropriate categories by themselves, computing agreement on these articles was not straightforward. 

In essence, \textbf{six}\footnote{This is the value of K chosen for our experiments in later sections.} sub-categories for the ``article/editorial" OCR category were found by human annotators and are illustrated in Table~\ref{subCat}. It must be noted that in this application, it is hard to obtain ``ground truth" or a ``gold standard" which can be used for further labeling. Consequently, we are forced to rely on the subjective opinion of annotators who sometimes disagree on labels. The six categories described above are also referred to as \emph{inferred ground truth labels} in later sections of this paper. There is considerable interest in the research community on whether this subjective labeling at low cost is indeed useful for machine learning algorithms \cite{Raykar_09,Hsueh_09}.
\begin{table}
\tbl{Sub-Categories formed by humans in the random sample of 25 articles.\label{subCat}}
{
\begin{tabular}{r c c}
\hline
ID & Category & Article counts \\\hline
1 & politics, elections, governmental appointments & 7\\
2 & medicine, public health and safety & 1\\ 
3 & death & 3 \\ 
4 & arts, human interest, entertainment & 2 \\ 
5 & marriage & 1 \\ 
6 & Other & 11 \\ \hline
\end{tabular}
}
%\caption{Sub-Categories formed by humans in the random sample of 25 articles.}
%\label{subCat}
\end{table}

The interview section of the annotation task indicated that small or singleton categories lead to less agreement among humans; these outliers do not fit into a larger category easily and this raised confusion and difficulty in categorization. Thus, learning from more examples of similar kind was the norm.

Many of the annotators based the initial decision of the number of categories by reading the headlines of the articles and making notes; a feedback loop was almost always involved where annotators refined the initial estimates based on more careful and thorough reading of the articles. Finally, since it was not clearly indicated whether an article is allowed to belong to multiple categories, this question was raised by several annotators. The time recorded by annotators indicates that it took anywhere between 45 mins - 2 hrs to categorize all the $25$ articles.

%\section{Refining Parameters of the K-Means Algorithm}
\section{Incorporating Prior Knowledge into Clustering Algorithms}
\label{unsupervised}
The information provided by annotators, albeit subjective, provides important insights about how documents in the archive can be grouped together. Thus, studying whether this \emph{domain knowledge} can be incorporated into automated document clustering algorithms would be beneficial. 
 
Clustering (or unsupervised learning) is ubiquitously used in machine learning problems where labels are not easily available or are difficult to generate. Given a data set $X$, a \emph{partitional} clustering algorithm \emph{groups} the data into $K$ block sets  and thus provides structure to previously unstructured data. This is particularly useful in the context of the NYPL historic newspaper archive since the documents in that repository have no prior labels other than the broad categorization provided by the OCR software which can be inaccurate. A clustering algorithm can thus be used to generate a taxonomy amongst similar articles. 

A wide variety of clustering algorithms exist including the $K$-means algorithm, hierarchical clustering, the expectation maximization algorithm and their variants to name a few popular algorithms. In this study we focus on the $K$-means algorithm which requires setting a large number of parameters such as the number of clusters, an appropriate distance function and a mechanism to initialize centroids. To estimate the performance of the algorithm, it is common practice to compare the labels obtained from it with the ``ground truth". However, ``ground truth" may not be available and/or can be subjective. 
%This section studies mechanisms of incorporating prior knowledge from multiple annotators into the $K$-means algorithm and uses this information to evaluate annotators.  

Semi-supervised clustering algorithms have become very popular in data mining and knowledge discovery \cite{Zhu_09a}. These algorithms are typically used in scenarios where only a small amount of data with prior knowledge (either as labels, constraints, etc.) is available in addition to a large proportion of unlabeled data. 
%If labeled data has representatives from all the clusters, then the algorithm can be easily used for the task of categorization (see surveys CITE). However, in some cases this knowledge is incomplete. Also 
The design of a semi-supervised clustering algorithm depends on the mechanism by which the prior knowledge is incorporated -- for example, (1) it may be available as \emph{pairwise constraints} implying there is pre-existing knowledge about whether two instances should belong to the same cluster (Must-link) or different clusters (Cannot-link); (2) labels associated with each instance or (3) inferring clustering constraints based on neighborhoods derived from labeled examples. The literature in semi-supervised learning however does not consider exhaustively scenarios where the labels provided are subjective. This is the focus of our research. 

In this paper, we first discuss the $K$-means algorithm with seeding as an example of how differences in parameterization of the clustering algorithms can affect the final results of the document clustering task. Next, we describe a different setting for the same document clustering task where domain knowledge comes as pairwise constraints. In either case the task of evaluation of the algorithms is hard, due to the subjective labels provided by our annotators. In the following subsections, we describe one after the other the standard $K$-means algorithm, its semi-supervised counterpart obtained by careful seeding and constrained $K$-means clustering with Must-link and Cannot-link constraints.

\subsection{The $K$-Means and Seeded $K$-means Algorithms}
\label{kmeans}
 One of the oldest and most commonly used clustering algorithms is the $K$-means algorithm \cite{Lloyd_57}, \cite{MacQueen_67}. Assume we are given an integer $K$ and a set of $N$ data points $X \subset \mathbb{R}^d$; the goal is to partition $X$ into $K$ clusters, $K < N$.  This can be achieved by choosing $K$ centroids $C_1, C_2, \cdots, C_K$ so as to minimize the potential function $\phi = \sum_{x \in X} \text{min}_{c \in C} \text{\emph{Dist}} [x - c]$, where $\text{\emph{Dist}}$ represents a distance function (such as squared euclidean, L1 norm). The basic steps of the algorithm are as follows: Arbitrarily choose initial $K$ centroids 
$C_1, C_2, \cdots, C_K$ from $X$; for each $i \in \{1, 2, \cdots K\}$ set the cluster $\mathcal{C}_i$ to be the set of all points in $X$ that are closer to centroid $C_i$ than they are to centroid $C_j,  \forall j \ne i$; for each $i \in \{1, 2, \cdots K\}$ set the cluster centroid $C_i = \frac{1}{|\mathcal{C}_i|}\sum_{x \in \mathcal{C}_i} x$; the last two steps are repeated until the process stabilizes and there are no new cluster assignments.

\subsubsection{Choice of Parameters:} There is much debate on how to choose a suitable number of clusters (\textbf{$K$}) appropriate for the data set. For our experiments we relied on human annotators to come to a consensus regarding the choice of an appropriate \textbf{$K$} as described in Section~\ref{pilot}. The other parameter that warrants some discussion is the choice of initial seeds; we have used two different seeding mechanisms in our experiments: (a) randomly chosen seeds which do not use information about clusters that humans produced (b) a semi-supervised K-Means algorithm called \textbf{Seeded K-Means} \cite{Basu_02}. This algorithm assumes that there exists $S \subseteq X $, called the ``seed set" on which \textit{supervision} is provided by annotators; thus, for each $x_i \in S$ the annotator indicates which cluster it seeds; there is at least one seed point $x_i$ per cluster.  
Once appropriate parameters have been set, the labels from K-Means are compared with those \textit{inferred} as ``ground-truth" in our pilot study. Note that all articles where annotators did not agree on labels were designated to a category called ``Other".
 
\subsubsection{Testing the validity of clusters:} 
\label{validity}
In order to measure the quality of the clusters produced by the K-means algorithm, we first compare them to human annotated data marking each instance as one of the six categories illustrated in Table~\ref{subCat}. This procedure allows us to quantitatively evaluate the system. The external cluster-validity measure used in this work was first suggested by Dom \cite{Dom_01} and is equivalent to mutual information when cluster labels and class labels are exactly the same. Let each data set $D$ have $n$ instances $O_1, O_2, \cdots, O_n$ and we want to partition it into $K$ clusters. Let $K=\{1,2, \cdots 6\}$ be the set of cluster labels and $C=\{1, 2, \cdots, 6\}$ be the expert annotated class labels assigned to the objects in $D$. Consider a two-dimensional contingency table, $\mathcal{H} = h(c, k)$ where $h(c,k)$ represents the number of objects labeled class $c$ are assigned to cluster $k$ by the algorithm. Then, if there is a perfect clustering $\mathcal{H}$ is a square matrix with only one non-zero element per row / column. The marginals are defined as $h(c) = \sum_k h(c,k)$ and $h(k) = \sum_c h(c,k)$. Since in our experiments the number of clusters are known apriori, the cluster-validity measure is essentially the empirical mutual information $\hat{I}(C,K) = \hat{H}(C) - \hat{H}(C|K)$, where $\hat{H}(C)= - \sum_{c=1}^{|C|} \frac{h(c)}{n} log \frac{h(c)}{n}$ and $\hat{H}(C|K)= -\sum_{c=1}^{|C|} \sum_{k=1}^{|K|} \frac{h(c,k)}{n} log \frac{h(c,k)}{h(k)}$.

In the second experiment, we questioned whether the choice of six categories was appropriate and instead compared the performance of seeded $K$-means with the standard $K$-means algorithm, setting the number of clusters as suggested by each annotator. As before, mutual information was recorded over multiple trials and the averaged results are presented in the following section.

\subsubsection{Empirical Evaluation:} The pilot study includes $25$ articles. A bag-of-words representation and tf-idf weights are obtained for these articles. Each article has 3210 features and one of possible six labels as indicated in section~\ref{interpret}. The K-Means algorithm with $K=6$ is run over 10 trials using both the random seeding and semi-supervised seeding. For semi-supervised seeding, one representative article from each category provided by human annotators is randomly selected for creating the seed; however care is taken to ensure that all six categories are represented by at least one seed. In each trial, the labels obtained after clustering are tested against the inferred ``ground-truth" generated by annotators (as described in section~\ref{interpret}) and mutual information is recorded. The average and standard deviation of mutual information obtained over all trials is presented in Table~\ref{mI}. Clearly, using Seeded K-Means with semi-supervision from annotators is more robust than the random seeding mechanism since the mutual information is higher and has a lower standard deviation over all the trials.

\begin{table}
\tbl{Mutual Information over 10 trials using Random vs Semi-supervised Seeding techniques. The inferred ``ground truth" sets $K=6$. \label{mI}}
{
\begin{tabular}{r c c}
\hline
%ID & Random Seeding & Seeded KMeans \\ \hline
%Trial 1 & 0.1849 & 0.2613 \\
%Trial 2 & 0.3501 &	0.2296 \\ 
%Trial 3	& 0.2783 &	0.173 \\
%Trial 4	& 0.1911 &	0.2613 \\ 
%Trial 5	& 0.1384 &	0.271 \\
%Trial 6	& 0.1978 &	0.4413 \\
%Trial 7	& 0.0492 &	0.1946 \\
%Trial 8	& 0.095	 & 0.1951 \\
%Trial 9	& 0.1203 & 	0.2823 \\
%Trial 10 & 	0.3377 & 0.273 \\
Seeding Algorithm & Mean & Std over 10 trials \\ \hline
Random Sampling & 0.19 &	 $\pm$0.10	\\
Semi-supervised with Seeding & 0.26 &  $\pm$0.07 \\ \hline
\end{tabular}
}
%\caption{Mutual Information over 10 trials using Random vs Semi-supervised Seeding techniques.}
%\label{mI}
\end{table}  

In the second experiment, the standard and seeded $K$-means algorithms were run with different values of $K$ as reflected in the annotator choices. The results are shown in Table~\ref{mIAnn}. For all the annotators, except annotator $3$, Seeded $K$-means performs better than the standard $K$-means algorithm. Closer investigation revealed that annotator $3$ had indeed provided multiple labels for a given article; for example, s/he surmised that a particular article in the pilot study could belong to the category ``arts/human interest/politics"; since the focus of our work was not on studying the impact of multiple labels, we decided to resolve ties by randomly selecting one label from all the suggestions; this process may have introduced bias and consequently affected the performance of seeded $K$-means. A better way to deal with multiple labelings would be to associate a probability of belonging to a particular class and then use this information to guide the seeding algorithm. 
\begin{table}
\tbl{Average Mutual Information and standard deviation over 5 trials using Standard and Seeded $K$-means algorithms\label{mIAnn}. The number of clusters is as proposed by each annotator. Ann $1 \cdots 6$ refers to Annotator 1 through 6 who participated in the pilot study.}
{
\begin{tabular}{ c c c c c c c }
\hline
%ID & Random Seeding & Seeded KMeans \\ \hline
 & Ann1 & Ann2 & Ann3 & Ann4 & Ann5 & Ann6 \\ \hline
No of clusters & 8 & 14 & 13 & 9 & 10 & 9 \\ \hline
Seeded $K$-Means Algorithm & 0.235$\pm$0.056 & 0.131$\pm$0.048 & 0.097$\pm$0.058 & 0.214$\pm$0.107 & 0.130$\pm$0.064 & 0.210$\pm$0.089	\\ \hline
$K$-means Algorithm  & 0.183$\pm$0.088 & 0.130$\pm$0.057 & 0.134$\pm$0.063 & 0.114$\pm$0.049 &0.075$\pm$0.042 & 0.097$\pm$0.062 \\ \hline
\end{tabular}
}
%\caption{Mutual Information over 10 trials using Random vs Semi-supervised Seeding techniques.}
%\label{mI}
\end{table}

In another experiment, we used the results from the pilot study to annotate unlabeled articles. We applied the Seeded K-Means algorithm with seeds suggested by annotators, on the remaining articles of the November 2nd, 1894 issue of \textit{The Sun} newspaper that were not included in the pilot study. At least one representative article from each category was randomly selected from clusters found by humans for creating the seed and care is taken to ensure that all six categories are represented. We ran the Seeded K-Means algorithm ten times on the unlabeled articles. For each run, the number of clusters is fixed at six, the cosine distance metric is used to compare similarity between instances, and the same technique (randomly choose one of the representative documents of a category as the centroid) is used for generating seeds. The labels obtained from each run can be considered as produced by an \textbf{automated annotator}. Since each automated annotator only provides labels between 1 and 6 we are able to use Krippendorff's alpha\footnote{We used the implementation available from \texttt{http://ron.artstein.org/resources/}} to measure inter-annotator agreement between them. It is seen that there is a very low agreement ($\alpha$=0.316) when 200 resamplings are used for calculating two-tailed $1\%$ confidence intervals. To illustrate this point further, we closely examined the labels provided by two representative automated annotators as shown in the confusion matrix illustrated in Table~\ref{conMat}. For these two automated annotators, there is complete agreement on sub-categories for $20.7\%$ of the articles used for blind testing; $61.9\%$ of articles labeled ``death" and $33\%$ of articles labeled ``Medicine" are correctly labeled. While these results are encouraging, there seems to be confusion in distinguishing between the ``election" and ``human interest" categories. It is worthwhile to note that humans also found it difficult to assign articles to the ``human interest" category and thus this task appears to be significantly harder. An interesting direction for future work is to use other mechanisms of finding representative seeds to be used with the Seeded K-Means algorithm. One such approach is to identify a centroid of the human clusters by calculating the cosine distance of each pair of documents in each human cluster, estimate the mean and then find the document closest to the mean as the seed.

\begin{table*}
\begin{center}
\tbl{Confusion Matrix generated by two runs of Seeded K-Means on blind test data formed by articles of the newspaper not considered for the pilot study.\label{conMat}}
{
\begin{tabular}{c | c c c c c c |c}
\hline
 & Elections & Medicine & Other & Death & Human Interest & Marriage & Total \\ \hline
Elections & 0 & 1 & 2 & 0 & 23 & 0 & 26\\
Medicine & 6 & 7 & 0 & 3 & 1 & 4 & 21\\
Other & 1 & 4  & 4&	2&	2&	9 & 22\\
Death & 7&	0&	0&	13&	0&	1 & 21\\
Human Interest &  2 &	16 & 	0&	5 &	1 &	1 & 25 \\
Marriage & 3&	0&	14&	0&	0&	3& 20\\ \hline
\end{tabular}
}
%\caption{Confusion Matrix generated by two runs of Seeded K-Means on blind test data formed by articles of the newspaper not considered for the pilot study.}
%\label{conMat}
\end{center}
\end{table*}  
 
\subsection{Constrained Clustering} 

The simplest form of constrained clustering uses instance-level constraints, introduced by Wagstaff et. al. \cite{Wagstaff_01}. There are primarily two types of constraints -- \emph{must-link} and \emph{cannot-link}.  A must-link constraint, defined as $c_=(a,b)$, requires a pair of points $a$ and $b$ to appear in the same cluster. It is an equivalence relation, hence it is symmetrical, reflexive, and transitive. This implies, if $c_=(a,b)$ and $c_=(b,c)$, then $c_=(a,c)$.  A cannot-link constraint, written as $c_{\ne}(a,b)$, on the other hand, restricts both points from being part of the same cluster.

%[PROVIDE ILLUSTRATIVE EXAMPLES]
An example of the above from our study is the following: assume an annotator has classified articles $1, 3, 6, 7$ in one category, and articles $2, 4, 5$ in another category. The constrained clustering algorithm would create a series of must-links which might include $c_=(1,3)$, which implies articles 1 and 3 are required to be part of the same group.  It would also generate a set of cannot-links that may include $c_{\neq}(3, 4)$, meaning articles 3 and 4 cannot be part of the same group.

%These instance-level constraints seem very simple, but they are extremely powerful and can potentially partition the data set $X$ and classify objects as

%If a clustering algorithm is given enough constraints, it can partition any domain $X$.  For our purposes, this means the clusterer would be able to accurately classify articles in any way a human annotator would.

%<<<<<< .mine
%=======
Constrained clustering can be categorized as (1) constraint-based and (2) distance-based. In constraint-based clustering, the algorithm utilizes constraints only after it has already classified all the points into initial clusters such as by using the $K$-means algorithm. It then verifies if there are any constraint violations and reallocates points to resolve violations. There can be a strict or soft enforcement of constraints. In strict-enforcement, no instance can violate any constraint. It either outputs a solution, or fails. A soft-enforcement allows violations, but adds penalties\footnote{Penalties can be distance metrics or conditional probabilities.} when required.  The clustering algorithm seeks the best feasible assignments and always produces a solution. Distance-based constrained clustering, on the other hand, treats constraints as weights on the distance functions. A must-link ``shrinks" the distance between two instances, while a cannot-link ``widens" the distance between them. 

The empirical results presented in this paper use a constraint-based algorithm (Pairwise Constrained Clustering K-Means (PCKMeans) \cite{Basu_04}) with probabilistic penalties. The algorithm is presented in Figure \ref{pckmeansalgo}. As an initialization step, it receives a list of constraints and generates the transitive closure of the must-links. It is important to note that this step makes it susceptible to noise. Following this, the standard K-means algorithm is run. The difference between the standard K-means and PCKmeans occurs in terms of when and how they exploit the constraints.

\begin{algorithm}
	\SetKwData{Left}{left}\SetKwData{This}{this}\SetKwData{Up}{up}
	\SetKwFunction{Union}{Union}\SetKwFunction{FindCompress}{FindCompress}
	\SetKwInOut{Input}{input}\SetKwInOut{Output}{output}
	\SetKw{Assign}{assign cluster:} \SetKw{Estimate}{estimate means:}
	\SetKwBlock{Method}{method}{}
	\Input{Set of data points $\chi = \{x_i\}^n_{x=1}$, set of must-link constraints $M = {(x_i, x_j)}$, set of cannot-link constraints $C = {(x_i, x_j)}$, number of clusters $k$, weight of constraints $w$.}
	\BlankLine

	\Output{Disjoint $k$ partitioning $\{\chi_h\}^k_{h=1}$ of $\chi$ such that objective function $\tau_{pckm}$ is (locally) minimized.}
	\BlankLine \BlankLine

	\Method
	{
		1. Initialize clusters.\\
		\Indp
			Create the $\lambda$ neighborhoods $\{N_p\}^\lambda_{p=1}$ from $M$ and $C$.\\
			Sort the indices $p$ in decreasing size of $N_p$.\\
		\Indm
		\uIf{$\lambda \geq k$}   {Initialize $\{\mu^{(0)}_h\}^k_{h=1}$ with centroids of $\{N_p\}^k_{p=1}$.}
		\uElse
		{
			Initialize $\{\mu^{(0)}_h\}^k_{h=1}$ with centroids of $\{N_p\}^\lambda_{p=1}$. \\
			\uIf{$\exists$ point $x$ cannot-linked to all neighborhoods $\{N_p\}^\lambda_{p=1}$.} {initialize $\mu^{(0)}_{\lambda+1}$ with x.}\\
			Initialize remaining clusters at random.\\
		 }
		\BlankLine
		2. Repeat until convergence.\\
		\Indp
			\Assign{Assign each data point x to the cluster $h^*$ (i.e. set $\chi^{(t+1)}_{h^*}$), for $h^* = argmin_h(\frac{1}{2}||x - \mu^{(t)}_h||^2 + w\sum_{(x,x_j) \in M}1[h \neq l_j] + w\sum_{(x,x_j) \in C}1[h = l_j])$.} \\
			\Estimate{$\{\mu^{(t+1)}_h\}^k_{h=1} \leftarrow \{\frac{1}{|\chi^{(t+1)}_h|}\sum_{x \in \chi^{(t+1)}_h}x\}^k_{h=1}$.} \\
			$t \leftarrow (t + 1)$.\\
	}
	\caption{PCKMeans Algorithm}
	\label{pckmeansalgo}
\end{algorithm}
%\DecMargin{1em}

\subsubsection{Effectiveness of Constraints:}  
How can we tell how useful constraints really are to the clustering process?  This is important to consider because it may be possible that constraints introduced hurt instead of improving performance \cite{Davidson_06}. The effectiveness of a set of constraints on a clustering problem can be measured by \emph{informativeness} and \emph{coherence}. \\

\noindent \textbf{Informativeness} is a measure of how much additional information about the domain the constraints provide to the clustering algorithm that it was not able to determine on its own. It is defined as:

\vspace{5mm}
\noindent
\( I_A(C) = \frac{1}{|C|}[\sum_{c \in C} unsat(c,P_A)] \)

\vspace{5mm}
\noindent
where $C$ is the set of constraints, $A$ is an unconstrained clustering algorithm, $P_A$ is the unconstrained clustering results of running $A$, and $unsat(c,P_A)$ is 1 if $P_A$ does not satisfy $c$ and 0 otherwise. \\

%\begin{figure}
%	\begin{center}
%		\includegraphics[height=0.15\textheight]{projections.png}
%		\caption{Illustrations of the three possible cases when computing the overlap between two constraints $\vec{a}$ and $\vec{b}$.  Reprinted from \cite{Davidson_06} POSSIBLY REDO THIS FIGURE}
%		\label{projections}
%	\end{center}
%\end{figure}

\noindent \textbf{Coherence} tries to determine how ``contradictory" the information is. It calibrates the level of agreement between constraints in set $\mathcal{C}$, using a distance metric $\mathcal{D}$. 
%A must-link can be considered an ``attractive force" and a cannot-link a ``repulsive force". When viewed this way, the two types of links are opposing forces.  
It is measured by the projected overlap between two constraints i.e. how much overlap there is when one constraint is projected along the direction of the other. 
\begin{figure}
	\begin{center}
		\includegraphics[height=0.25\textheight]{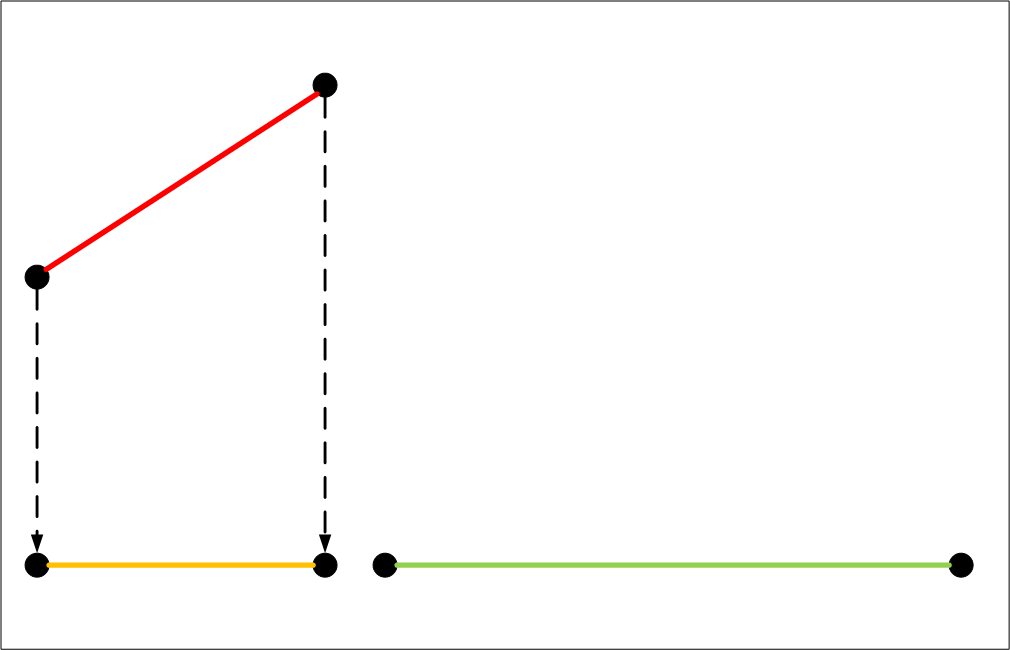}
		\caption{First illustration of projected overlap between a must-link and a cannot-link.  The must-link is the green line, the cannot-link is the red line, and the orange line is the projection of the cannot-link onto the must-link.  In this example, there is no overlap between the two links.}
		\label{overlapnone}
	\end{center}
\end{figure}

\begin{figure}
	\begin{center}
		\includegraphics[height=0.25\textheight]{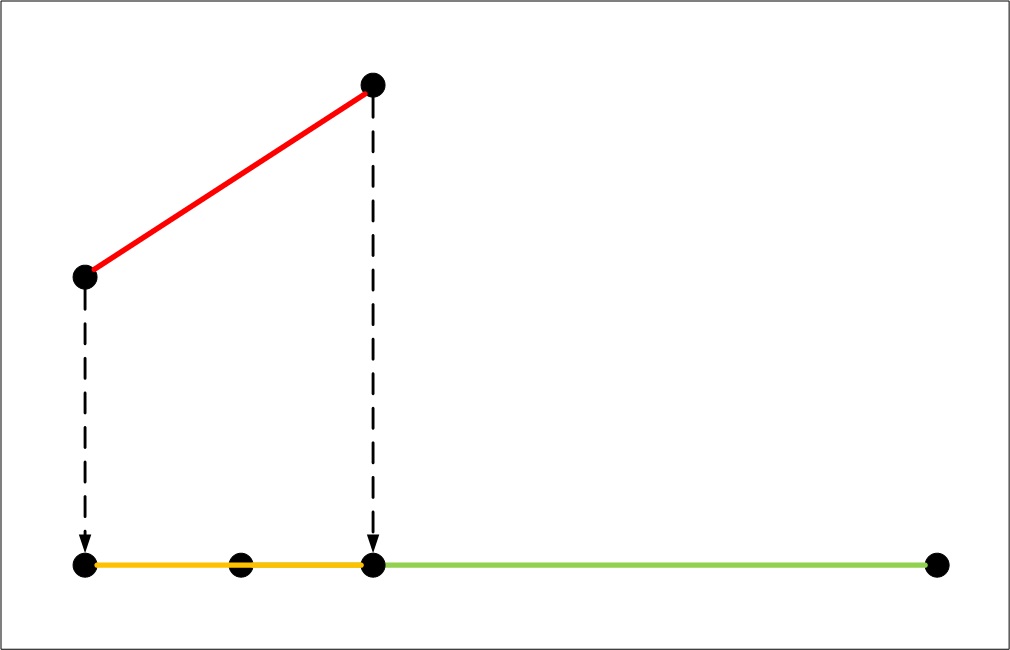}
		\caption{Second illustration of projected overlap between a must-link and a cannot-link.  The must-link is the green line, the cannot-link is the red line, and the orange line is the projection of the cannot-link onto the must-link.  In this example, there is some overlap between the two links.}
		\label{overlapsome}
	\end{center}
\end{figure}

\begin{figure}
	\begin{center}
		\includegraphics[height=0.25\textheight]{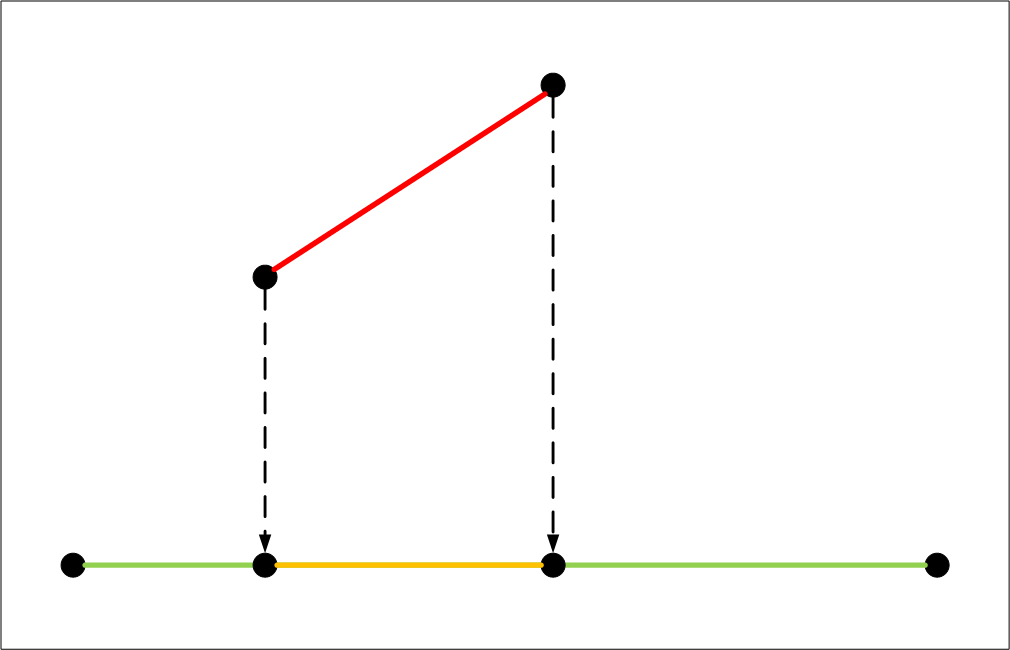}
		\caption{Third illustration of projected overlap between a must-link and a cannot-link.  The must-link is the green line, the cannot-link is the red line, and the orange line is the projection of the cannot-link onto the must-link.  In this example, there is complete overlap between the two links.}
		\label{overlapall}
	\end{center}
\end{figure}

%For example, consider four points, two of which (say $A$ and $B$) are close to each other and the other two points, $C$ and $D$, are fairly close to each other but relatively far from $A$ and $B$.  If we assume that there is a must-link between $A$ and $C$ and a cannot-link between $B$ and $D$, then these four points would be considered incoherent. On the one hand, the must-link implies that these four points belong in the same group, while the cannot-link indicates that $A$ and $B$ do not belong with $C$ and $D$. To determine the coherence of two constraints $a$ and $b$, the projected overlap of one on the other is estimated. 

\noindent Let $\vec{a}$ be a cannot-link and $\vec{b}$ a must-link, then projection is estimated as follows:
%as illustrated in Figure \ref{projections},
% as follows:

%\vspace{5mm}
%\noindent
\begin{center}
\( \vec{p} = proj_{\text{ }\vec{b}\text{ }} \vec{a} =  (\vert{\vec{a}} \vert \text{ } cos \theta) \text{ } \frac{\vec{b}}{\vert{\vec{b}} \vert}  \)
\end{center}
\noindent where $\theta$ is the angle between the two vectors. To calculate how much of the projection of $\vec{a}$ overlaps with $\vec{b}$, we compute the distance corresponding to the three cases: 

$$
overlap_b(a) = \left\{ \begin{array}{rl}
  0 &\mbox{ if $d_{b_2,b_1} \leq d_{b_2,p_2}, d_{b_2,b_1} \leq d_{b_2,p_1}$ } \\
  d_{b_1,p_2} &\mbox{ if $d_{b_2,p_2} < d_{b_2,b_1}, d_{b_2,p_1} \geq d_{b_2,b_1}$} \\
  d_{p_1,p_2} &\mbox{ if $d_{b_2,p_2} < d_{b_2,b_1}, d_{b_2,p_1} < d_{b_2,b_1}$}
       \end{array} \right.
$$

%\begin{equation}
%\noindent
%\begin{center} 
%\begin{eqnarray*}
%overlap_b(a) &=&  0, &  \text{if } d_{b_2,b_1} \leq d_{b_2,p_2}, d_{b_2,b_1} \leq d_{b_2,p_1} \\
%	    &=& d_{b_1,p_2}, & \text{if  } d_{b_2,p_2} < d_{b_2,b_1}, d_{b_2,p_1} \geq d_{b_2,b_1} \\
%	    &=& d_{p_1,p_2}, & \text{if }  d_{b_2,p_2} < d_{b_2,b_1}, d_{b_2,p_1} < d_{b_2,b_1}.
%\end{eqnarray*}

%\end{center}
%\end{equation}
%\vspace{5mm}
\noindent where $b_1, b_2, p_1, p_2$ are the beginning and end co-ordinates of vector $\vec{b}$ and the projection of $\vec{a}$ given by $p$. 
Figures \ref{overlapnone}, \ref{overlapsome}, and \ref{overlapall} provide examples of the projection scheme. There are two ways by which constraints can have \emph{zero} projected overlap: (1) they are orthogonal to each other, so that neither link interferes with the other or (2) they are both the same type of links (both are must-links or both are cannot-links), so any overlap that exists does not matter. Coherence ($COH$) of a constraint set $\mathcal{C}$ using a distance metric $\mathcal{D}$ is then be defined as the fraction of ML-CL\footnote{ML: Must-Link, CL: Cannot-Link} constraint pairs in the constraint set $\mathcal{C}$, that have zero projected overlap i.e.

\begin{equation}
\label{coh}
%\begin{center}
%\vspace{5mm}
%\noindent
COH_{\mathcal{D}}(\mathcal{C}) = \frac{\sum_{m \in C_{ML} \text{ , } c \in C_{CL}} \delta(overlap_c\text{ m} = 0 \text{ and } overlap_m \text{ c} = 0)}{|C_{ML}| |C_{CL}|}
%\end{center}
\end{equation}
%\vspace{5mm}
\noindent where $C_{ML}$ and $C_{CL}$ represent the set of must-link and cannot-link constraints respectively in $\mathcal{C}$; $|C_{ML}|$ and $|C_{CL}|$ represents the cardinality of each set.
A clustering algorithm, when given a constraint set with low coherence, gets confused on how to properly label points.

\begin{figure}
	\begin{center}
		\includegraphics[height=0.25\textheight]{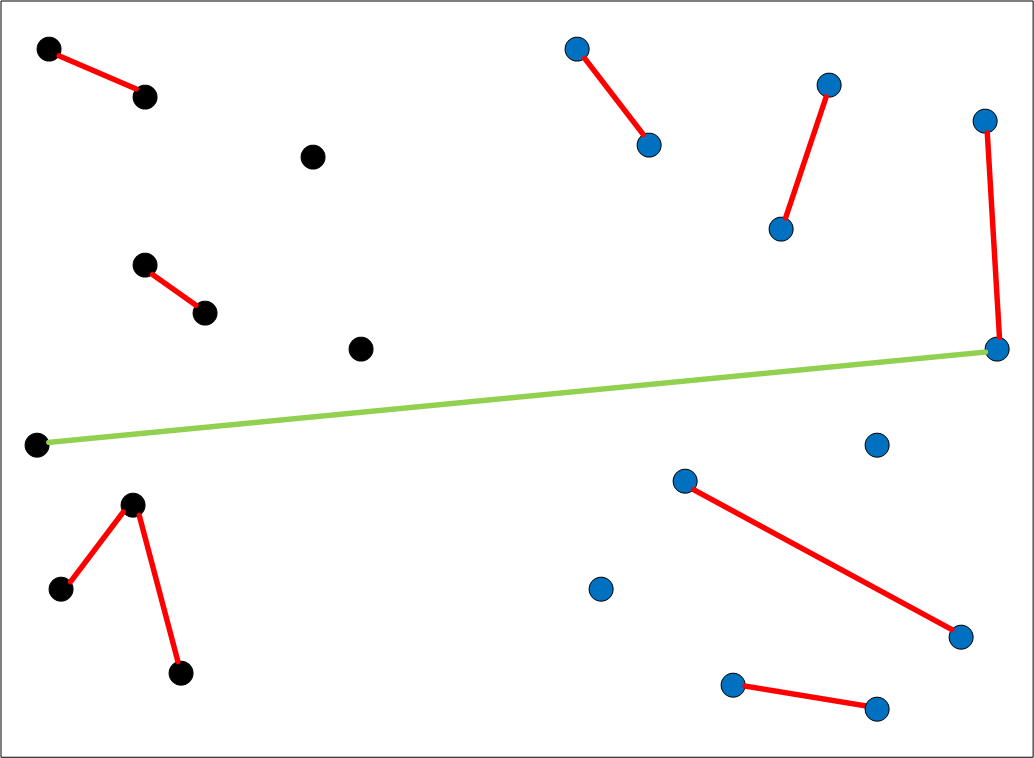}
		\caption{An example of a constraint set with high informativeness.  The points represent articles.  The black dots represent one cluster, and the blue dots represent a second cluster.  This is typical of how a simple KMeans algorithm would classify points based on distance.  The red edges are cannot-link constraints, and the green edges are must-link constraints.  This constraint set indicates that many of the points close together do not belong in the same set, while the two points that the must-link connects belong in the same set.  A simple KMeans algorithm would not partition the dataset in this way, meaning the constraints provide very valuable information.}
		\label{highinfo}
	\end{center}
\end{figure}

\begin{figure}
	\begin{center}
		\includegraphics[height=0.25\textheight]{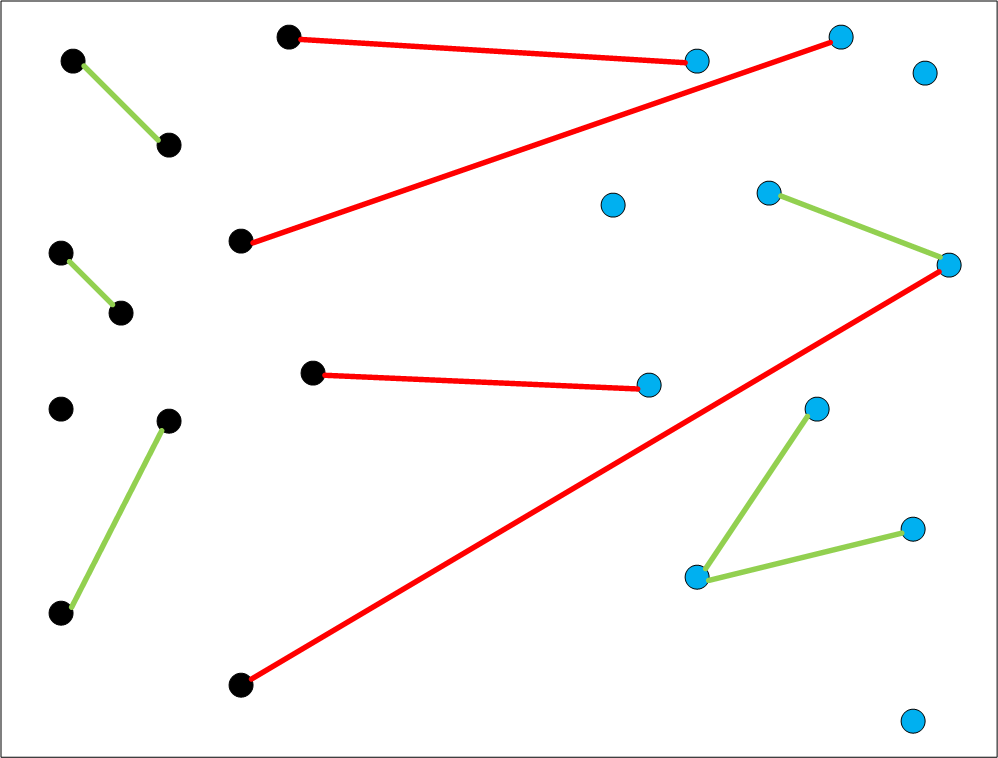}
		\caption{An example of a constraint set with low informativeness.  The points represent articles.  The black dots represent one cluster, and the blue dots represent a second cluster.  This is typical of how a simple KMeans algorithm would classify points based on distance.  The red edges are cannot-link constraints, and the green edges are must-link constraints.  This constraint set does not provide any additional information that conflicts with how the simple KMeans algorithm would classfiy these points.  All the links between a point in the first cluster and a point in the second cluster are cannot-links, while all the links connecting two points within the same cluster are must-links.}
		\label{lowinfo}
	\end{center}
\end{figure}

\begin{figure}
	\begin{center}
		\includegraphics[height=0.25\textheight]{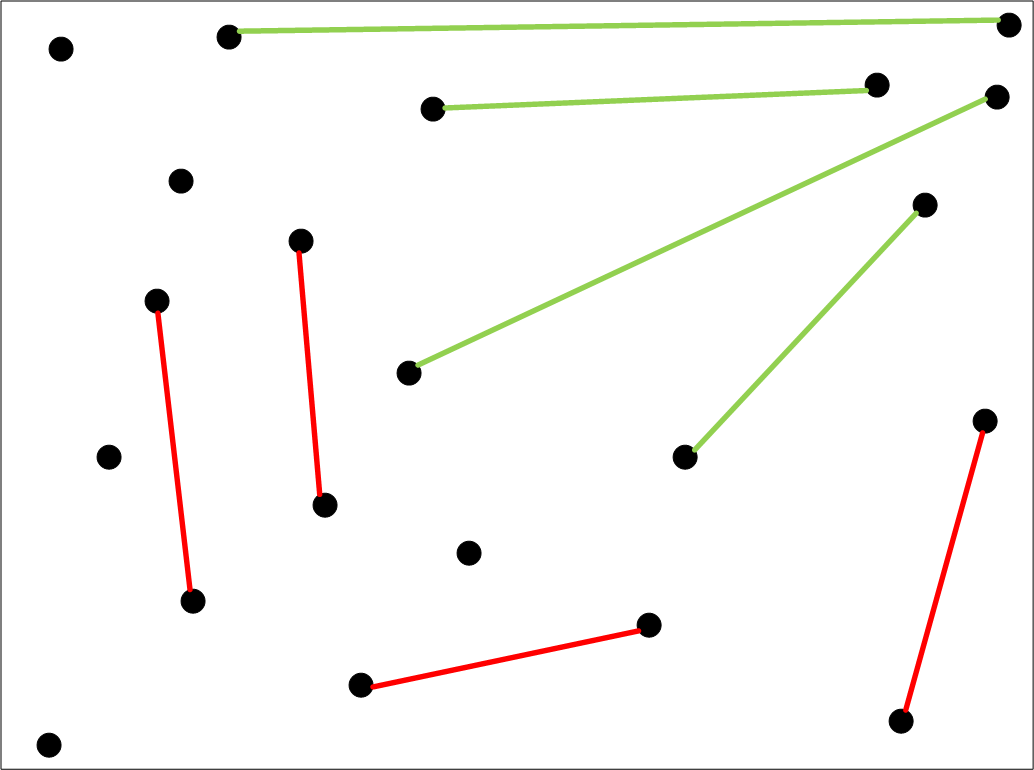}
		\caption{An example of a constraint set with high coherence.  The points represent articles.  The red edges are cannot-link constraints, and the green edges are must-link constraints. The points in the region near the two points connected by the constraint should have the same type of link. In this figure, all the must-links connect to points very near each other and do not show any contradictions.}
		\label{highcoh}
	\end{center}
\end{figure}

\begin{figure}
	\begin{center}
		\includegraphics[height=0.25\textheight]{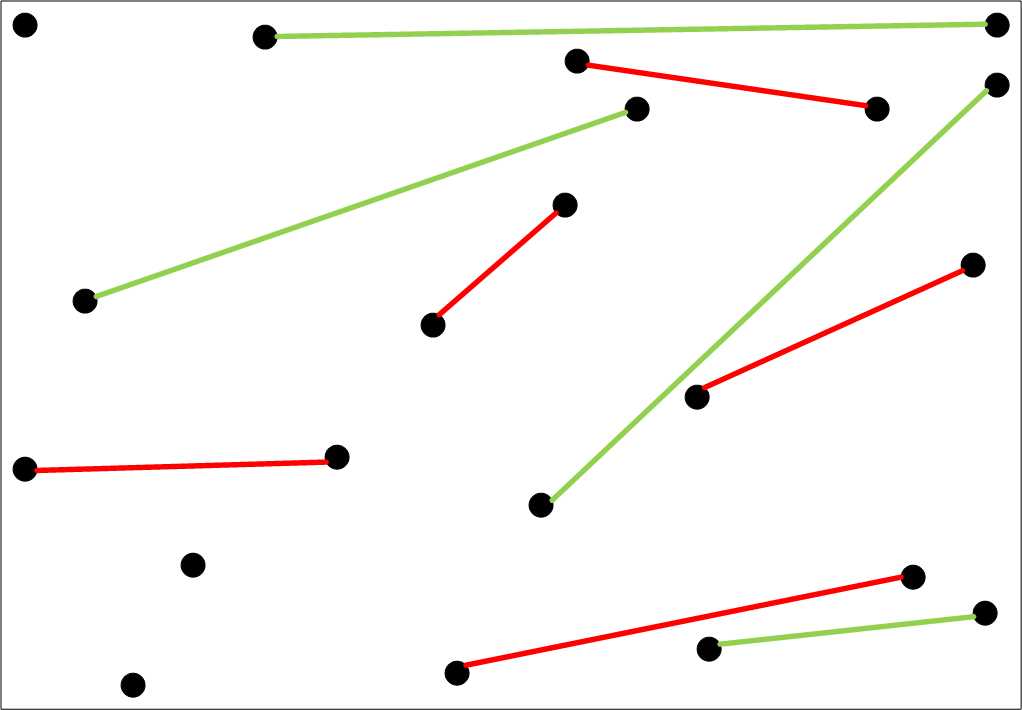}
		\caption{An example of a constraint set with low coherence.  The points represent articles.  The red edges are cannot-link constraints, and the green edges are must-link constraints. The links connecting the points in the upper right-hand corner with other points in the figure are a mixture of must-links and cannot-links that seem to contradict each other. This constraint set can confuse the PCKMeans algorithm.}
		\label{lowcoh}
	\end{center}
\end{figure}

Davidson et. al. \cite{Davidson_06} indicates that constraint sets with high informativeness and high coherence improve performance, while sets with low informativeness and low coherence hurt performance.  Low informativeness means the constraint set does not provide much helpful information.  Low coherence means the information provided by the constraints set is contradictory and confusing. This is further illustrated in Figures \ref{highinfo}, \ref{lowinfo}, \ref{highcoh} and \ref{lowcoh}.

%\noindent
%\textbf
\subsubsection{Modeling Annotators and Assessing their Quality}
\label{compareAnn}
Online digital archives such as the NYPL historic newspaper archive are unlabeled and getting a dataset labeled by annotators can be time consuming, if not impossible. With the advent of crowdsourcing, getting cheap labels is relatively easy. How good are these labels? People with different levels of expertise -- novices, scholars, biased and malicious annotators may provide inexpensive labels but chaffing meaningful information from it could be challenging. Accessing the ``quality" of labels is therefore of interest. 

%A clustering algorithm can group instances into different categories and act much like an ``automated" annotator. 

In the PCKMeans algorithm, the constraints can be used to encapsulate the prior knowledge an annotator has. The \emph{informativeness} of constraints then provides a qualitative measure of how good the suggested constraints are -- in other words, informativeness measures how much additional information about the domain these constraints have provided by comparing the same clustering algorithm  with and without\footnote{It should be noted that for the comparison to be meaningful, the same set of initial parameters for the unconstrained clustering algorithm should be used; for example if using the KMeans algorithm, the unconstrained algorithm should be trained each time with the same initial seeds, distance metric and the number of clusters.} constraints. An annotator with more ``informative" constraints is clearly preferred over one whose constraints do not provide additional information about the domain. 

For our experiments, we modeled each annotator by the PCKMeans clustering algorithm; different sets of constraints were generated from the labels provided by the annotator; the performance of the KMeans algorithm with and without constraints was measured using informativeness.

\begin{figure}[h]
	\begin{center}
		\includegraphics[height=0.38\textheight]{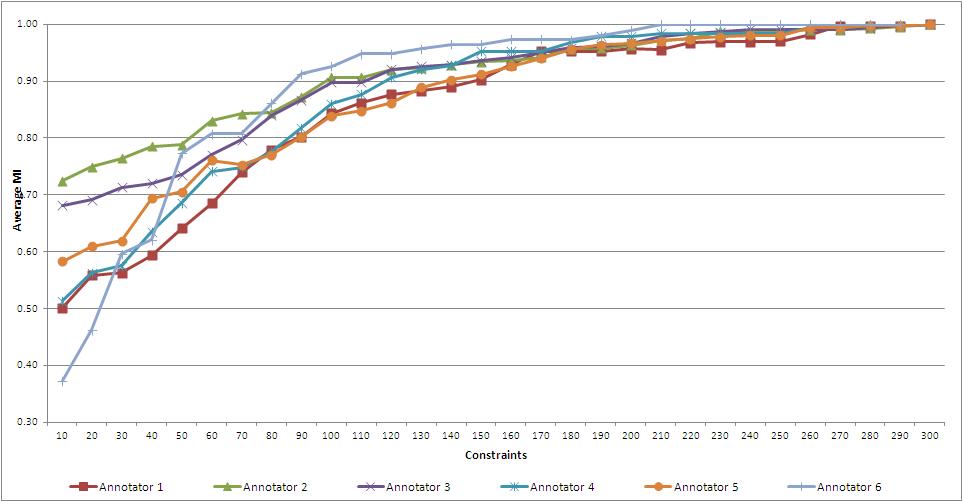}
		\caption{Average Mutual Information vs. Constraints.  The mutual information compares each annotator's PCKMeans results with his/her own labels across increasing number of constraints.  As expected, as the number of constraints increased, the PCKMeans results for each annotator became more in line with his/her labelings.}
		\label{fig:mil}
	\end{center}
\end{figure}

\begin{figure}
	\begin{center}
		\includegraphics[height=0.38\textheight]{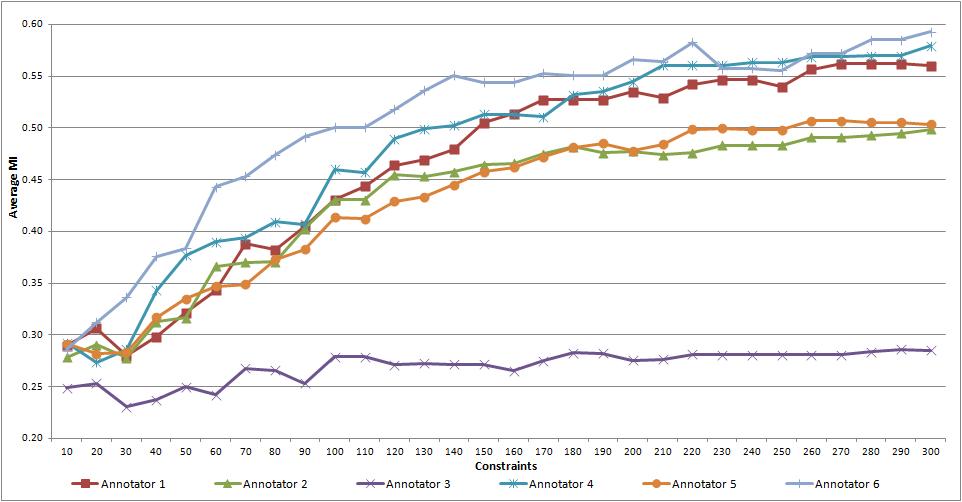}
		\caption{Average Mutual Information vs. Constraints.  Here, the mutual information compares each annotator's PCKMeans results with the inferred ground truth labels.  Here, we didn't expect the values to converge to one as the number of constraints increased, but we did expect the informativeness to improve.  The graph indicate that the level of informativeness depended on the difference between the number of clusters each annotator assigned and the number in the inferred ground truth.}
		\label{fig:mip}
	\end{center}
\end{figure}

\subsubsection{Empirical Evaluation:}  For our experiments, we used the WekaUT extension for Weka from \texttt{http://www.cs.utexas.edu/users/ml/risc/code/}. \\

\noindent \textbf{Experiment 1: Replicating annotator performance by PCKMeans} 

In the first experiment, we wanted to determine if PCKMeans could accurately replicate the performance of annotators and classify the twenty-five articles the  way each annotator did. We ran five trials of the clustering algorithm on the data for each annotator for different number of constraints, varying them from 10 to 300 in increments of 10. The constraints were generated as follows from the labels provided by the annotators: Randomly sample two instances;  if they were assigned the same class label by the annotator there is a MUST link between them; if they are assigned different class labels, they have a CANNOT link between them. Note that $C_2^{25} = 300$ was the maximum number of constraints that could be generated from the pilot study data. The mutual information (as described in Section~\ref{validity}) between each annotator's PCKMeans clustering results vs his/her own labels is estimated. The average mutual information over all the five trials is shown in Figure~\ref{fig:mil}. It appears that that PCKMeans can accurately replicate the annotators only when supplied with a sufficiently large number of constraints. The mutual information results reached one only after 200 constraints were supplied to the algorithm. In the second experiment, the mutual information between each annotator's PCKMeans results and those inferred from the pilot study were computed. The results are illustrated in Figure~\ref{fig:mip}. It was not expected that each annotator's labels would converge\footnote{Convergence to one meant that the annotator's label exactly matched the inferred ground truth label.}  to $1.0$ and indeed some annotators such as Annotator 2, 3 and 5 have low average mutual information. However, the informativeness improved as expected.
Another interesting thing was that the annotators who supplied the largest numbers of categories were the ones who had the lowest mutual information score since they were providing far more labels than what was inferred from ground truth. \\

\begin{figure}
	\begin{center}
		\includegraphics[height=0.38\textheight]{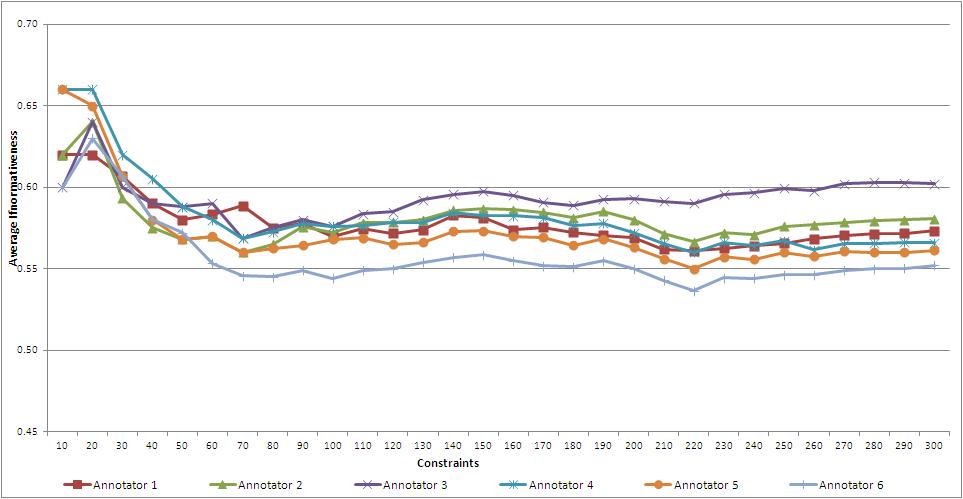}
		\caption{Average Informativeness vs. Constraints.  Here, the informativeness is measured by taking each annotator's PCKMeans results over the results of the standard KMeans for the inferred ground truth data.}
		\label{fig:ais}
	\end{center}
\end{figure}

\begin{figure}
	\begin{center}
		\includegraphics[height=0.38\textheight]{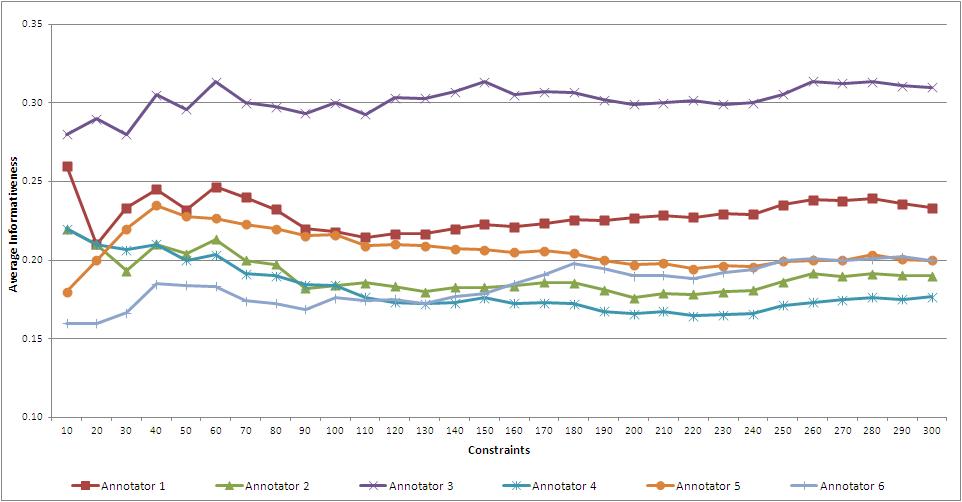}
		\caption{Average Informativeness vs. Constraints.  Here, the informativeness is measured by taking each annotator's PCKMeans results over the results of PCKMeans for the inferred ground truth data.}
		\label{fig:aip}
	\end{center}
\end{figure}
 
\noindent \textbf{Experiment 2: Comparing Annotators}

In the next experiment, we had six PCKMeans clusterers, one trained for each annotator; we compared the output of each PCKMeans clusterer with the results of running standard KMeans (Figure~\ref{fig:ais}) %and PCKMeans (Figure~\ref{fig:aip}) 
for inferred ground truth labels. This enabled us to measure the \emph{informativeness} of constraints for each PCKMeans clusterer and thereby provide a way to compare annotators as described in Section~\ref{compareAnn}. As before, the number of constraints was varied from 10 to 300 in increments of 10. Each colored line in Figure~\ref{fig:ais} refers to the average performance of an annotator over five different trials. It is interesting to see that all of the annotators seem to have a high informativeness when the number of constraints is below 60. When the number increases beyond 60, the informativeness becomes more or less constant -- in other words, increase in number of constraints arbitrarily does not provide more prior knowledge with respect to the unconstrained problem. It is also interesting to note that annotators 2 and 3 maintained a high informativeness compared to all the other annotators -- a closer look at their annotations revealed that they had provided the largest number of categories in the pilot study; not only were they thinking hierarchically when providing the categories, they also seemed much more detail-oriented in their approach to providing annotations. \\

\noindent \textbf{Experiment 3: Studying the impact of providing (and inferring) different number of categories}

Another interesting experiment that we conducted was to test the output of each of the PCKMeans clusterer with the results of running PCKMeans (Figure~\ref{fig:aip}) for inferred ground truth labels. This allowed us to study the impact of an annotator suggesting different number of classes than those inferred from the pilot study. For example, if a PCKMeans clusterer with 10 constraints was generated from 13 class labels suggested by an annotator was it necessarily better (in terms of informativeness) than a PCKMeans algorithm with 10 constraints generated from 6 class labels as inferred from the pilot study?  In this case, annotator 3 still consistently had a higher informativeness than others, but surprisingly annotator 2 who provided the maximum number of classes in the pilot study did not have an overall high informativeness. This could be attributed to the fact that the constraints are generated randomly -- it would be much more useful to study the case where the pairwise constraints are provided manually by the annotators. \\

\noindent \textbf{Experiment 4: Studying Coherence}

\begin{figure}[h]
	\begin{center}
		\includegraphics[scale=0.75]{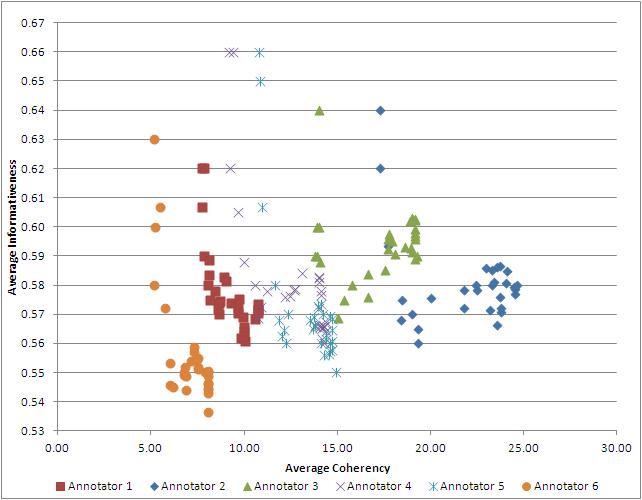}
		\caption{Average Informativeness vs. Coherence.  Here, the informativeness is measured by taking each annotator's PCKMeans results over the results of the standard KMeans for the inferred ground truth data.}
		\label{fig:cohs}
	\end{center}
\end{figure}

\begin{figure}[h]
	\begin{center}
		\includegraphics[scale=0.75]{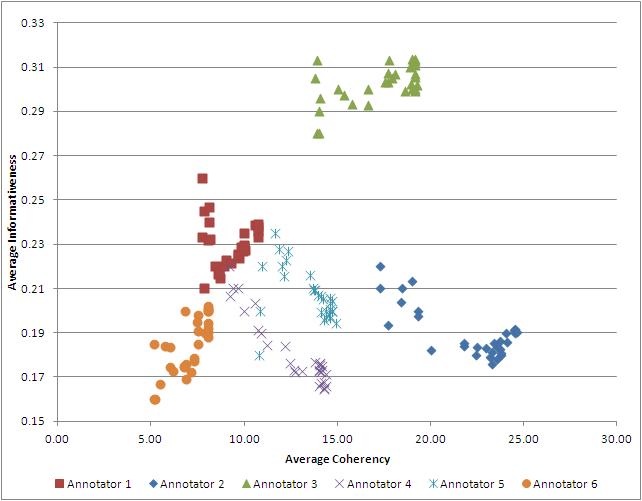}
		\caption{Average Informativeness vs. Coherence.  Here, the informativeness is measured by taking each annotator's PCKMeans results over the results of PCKMeans for the inferred ground truth data.}
		\label{fig:cohp}
	\end{center}
\end{figure}

Figures~\ref{fig:cohs} and ~\ref{fig:cohp} compare the informativeness against the coherence of the clusters generated in the two cases described above. Each dot in the figure represents the informativeness vs. coherence value for a given number of constraints\footnote{The reader is reminded that the constraints are varied from 10 to 300 in increments of 10 each.}. 

If we divide the graphs into four regions, each region indicates a different set of characteristics of the constraint set.  The upper right region, the area with high informativeness and high coherence, indicates that the annotator's labelings are extremely helpful to the clustering algorithm.  This is the ideal region.  The algorithm will very likely produce great results.  The lower left region of the graph indicates low informativeness and low coherence, meaning the classifications the annotator provides not only is not very helpful, but contradictory and confusing to the clusterer.  This is the worst region, and any results produced by PCKMeans algorithm will mostly be poor.  The other two regions are of mixed usefulness.  The upper left region means the annotator provides a lot of but confusing information, while the lower right region means the annotator provides clear but very little information.

\noindent As Figure~\ref{fig:cohp} shows, Annotator 3's datapoints fall closest to the ideal region, while Annotator 6's points fall in the bad region.  This indicates that Annotator 3 provides a better and clearer information to the clusterer.

\noindent In addition, it is helpful to look at how spread out the datapoints are. A correct clustering in one region means the characteristics of the annotator's labels are consistent across a varying number of constraints.  A wide and loose scattering means the labelings are not consistent.  Comparing Figures~\ref{fig:cohs} and ~\ref{fig:cohp}, one sees that the datapoints for all the annotators are clustered much more tightly when compared to the PCKMeans algorithm than when compared to simple KMeans.  This indicates that the PCKMeans ground truth is more consistent with what the annotators have in mind than the simple KMeans.

\begin{landscape}
\begin{figure}[h]
\begin{center}
\includegraphics[width=\columnwidth,height=0.9\textheight]{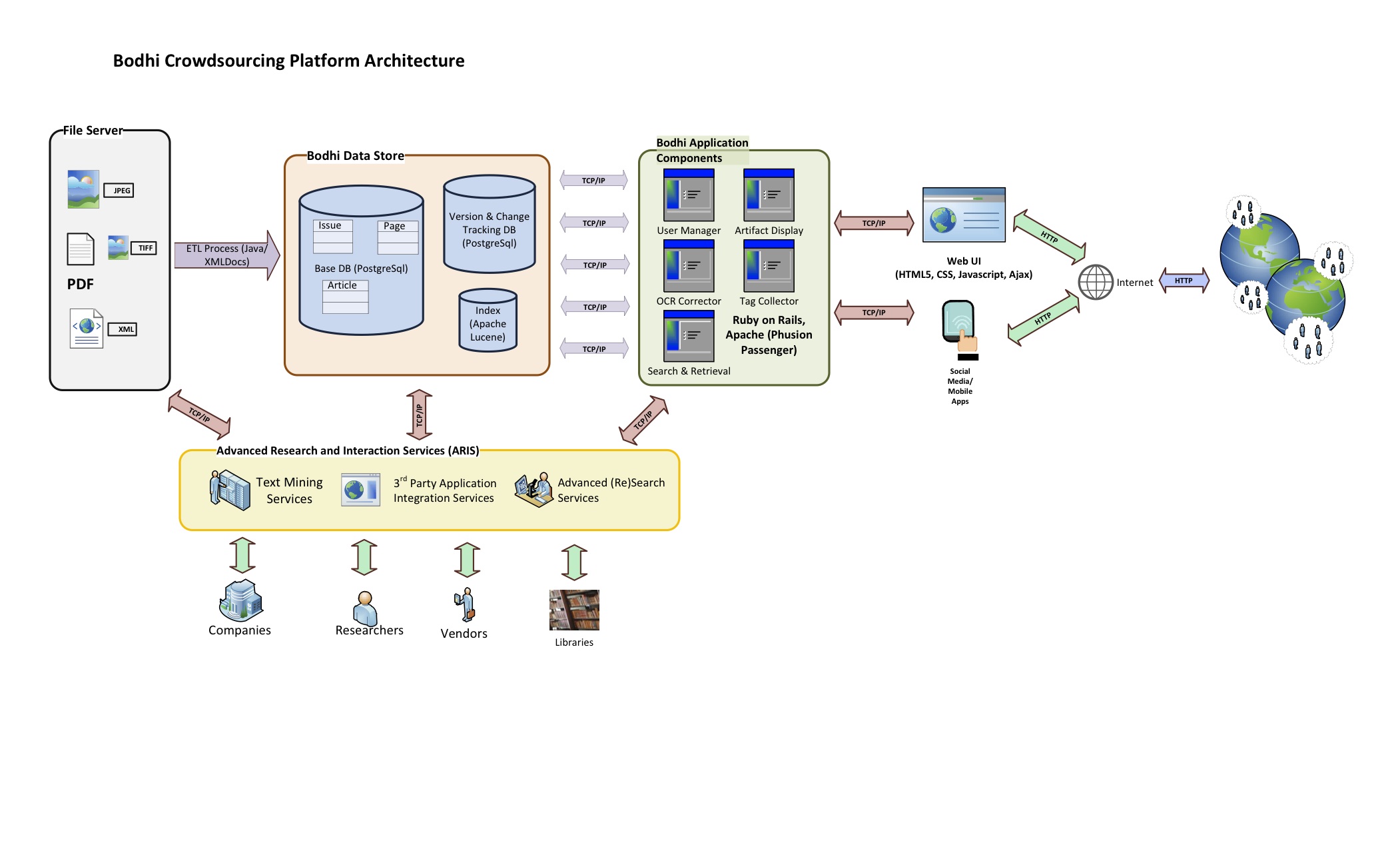}
\caption{Architecture and Implementation Details of the BODHI Crowd Sourcing System }
\label{bodhiArch}
\end{center}
\end{figure}
\end{landscape}

\section{The BODHI System: Large Scale OCR Correction and Tag Collection at the NYPL}
\label{tagworks}

To enable collaborative tagging, we are designing a system (named ``BODHI") that can be integrated with the current architecture used in National Digital Newspaper Program (NDNP) and allow patrons to correct garbled OCR text, enter keywords for tagging articles, provide useful information about segmentation of the article (for example -- is the article continued onto another page or not), links to other related articles, etc. These user-provided meta-data will augment the scanned text and image data obtained from the OCR scanning process. While the NDNP Content Management System has many modules to manage the newspaper digitization workflow from scanning microfilm to public delivery, the BODHI system focuses only on improving search and retrieval. 

%Figure~\ref{cArch} presents the current architecture of the NDNP and the proposed changes in BODHI are illustrated in Figure~\ref{pArch}. 

%With assistance from the NEH Digital Humanities Start-Up Grants, we have been able to build proof-of-concepts including: (a) Design and implementation of the OCR Correction and Tagging Interface also called the \emph{Article Manipulation} Module -- this includes an authentication module which enables only valid users of the system to make changes or provide tags. (b) Extract, Transform and Load text data in XML files into a PostgreSQL database (see Appendix for a detailed ER-Diagram (Figure~\ref{erd}) of the database). This is also the backend for the article manipulation module and will also contain user-generated meta-data. (c) Design of a small \emph{Pilot Study} involving six annotators reading a small sample of newspaper articles (d) Storing the annotations provided in the PostgreSQL database and (e) Using the annotations for a preliminary machine learning (text mining) study. 

Figure~\ref{bodhiArch} presents the architecture and digital technology that will be used in development of the ``BODHI" system. The file server stores the digitized newspaper images as jpeg, tiff, pdf or xml files (these are obtained from the NYPL after the OCR scanning process) -- this is subjected to Extract, Transform and Load (ETL) operations using software developed in Java and XML and stored into a PostgreSQL database. In addition to the storage of OCR data from the newspaper articles, the database is also capable of storing user registration information, version and change tracking as required (assuming the patrons of the archive may edit an article multiple times and add different annotations) and is indexed using Apache Lucene (\texttt{http://lucene.apache.org/java/docs/index.html}), which has an open-source Java-based indexing and search implementation as well as spellchecking, hit highlighting and advanced analysis/tokenization capabilities. The components of the BODHI application include a user manager, article display, OCR corrector, tag (or annotation) collector and search and retrieval manager. The prototype has been developed using Ruby-on-Rails and will be deployed using the Apache Phusion Passenger framework (\texttt{http://www.modrails.com/}). Once deployed on the library infrastructure, the web interface can be accessed by social media, mobile apps and the internet.

\section{Conclusion and Future Work}
\label{conc}
%[THIS NEEDS TO BE EDITED]
The New York Public Library has an archive of over 200,000 historical newspapers published between 1890 and 1920 which have been subjected to OCR and are currently stored in an online database making them accessible to patrons. Unfortunately search facilities on this database are rudimentary; newspapers are scanned on a page-by-page basis and article level segmentation is almost non-existent; the OCR scanning process introduces a lot of garbled text. In a bid to make these archives more accessible to the general public, text mining algorithms are being considered for categorization of articles. The OCR software provides a rough categorization, but a large chunk of the articles are labeled ``article/editorial" without division into further meaningful categories. Thus, articles dealing with medicine and crime are deemed to belong to the same category; this makes search and retrieval of articles difficult. We designed a pilot study to observe if humans were able to find coherent categories in a small subset of articles. Our results indicate that the presence of small and noisy clusters in the data made it difficult to find an agreement in the number of clusters. We also evaluated the quality of the annotation provided by humans by measuring how much additional information they could provide to help the clustering algorithm. More specifically, the \emph{informativeness} of constraints in a constrained clustering algorithm was used as a metric to evaluate how well they labeled articles. Our results from the pilot study are very encouraging and we are developing a large scale system (christened ``BODHI") in collaboration with the New York Public Library to collect tags and incorporate the ``wisdom of crowds" into machine learning algorithms. Future work involves development of more sophisticated non-parametric and bayesian text mining algorithms and experiments on a large scale using the deployed ``BODHI" system.

% Acknowledgments
\begin{acks}
This work is supported by funding from the National Endowment for Humanities, Grant No: NEH HD-51153-10. The authors would like to thank Dr. David Waltz for stimulating discussions during the initial phases of the project, Dr. Dragomir Radev for his generous and insightful comments on drafts of the paper, Sam Lee and Hatim Diab for help with infrastructure and system development.
\end{acks}

% Bibliography
\bibliographystyle{acmsmall}
\bibliography{flairs,tkdd,News,socialNet}

% History dates
\received{April 2012}{August 2012}{October 2012}

% Electronic Appendix
\elecappendix

\medskip

\section{User Interface Design of the BODHI System}
A prototype for the user interface in the BODHI system, which allows correction of OCR text and collection of tags from patrons was developed. The article manipulation module is capable of retrieving an article from the database based on a search criteria, displaying the scanned OCR text alongside a high resolution image, highlighting sections of it when clicked, allowing a user to edit the OCR text after checking the content in the high resolution image and storing the corrected text back into the database. Addition of tags and comments on an article-by-article basis is permissible. The module also has an user authentication mechanism for patrons registered to correct OCR, add notes or tags. 

Figure~\ref{scrShot} in the Appendix presents screen-shots from the prototype. The application utilizes Rails 3.1 (on Ruby 1.9.2), a Model-View-Controller framework to map rows from the PostgreSQL database into objects that may be utilized by the client. The viewer extends the functionality of a JQuery plugin, called ImgAreaSelect \\
(\texttt{http://odyniec.net/projects/imgareaselect/examples.html}), an open source independent project by Michael Wojciechowski. This plugin provides useful functions for manipulating an image, such as selection of a specific area, customization of the behavior of the selection box (the dotted box shown in Figure~\ref{scrShot}) and functions that expose key events (such as moving the selection box). In our application, the ImgAreaSelect plugin is utilized by loading two images -- one in low resolution and the other in high resolution. The low resolution image allows a user to select the part of the picture s/he wants enlarged. The high resolution image is displayed by the ``Viewer". Javascript code takes the selection of the low resolution image and scales to the a high resolution image and finally displays it in a box that overlaps the original image.

The backend for the article manipulation module stores precise $x$ and $y$ coordinates for each word -- the Rails application is capable of retrieving this data and using the ImgAreaSelect plugin to pinpoint a word that is clicked on both the low resolution preview picture and the magnified high resolution picture. The highlighting of a word is done by a simple JQuery call that manipulates the Cascading Style Sheet (CSS) attributes of the Document Object Model (DOM) element in the HTML.

A basic login system is implemented using Devise 1.5\footnote{\texttt{https://github.com/plataformatec/devise}}, an open source plugin ``gem" available to Rails. Each user will be required to create an account by specifying an email and password and use the information to actually correct a newspaper article and create tags.

\begin{figure}
\begin{center}
\includegraphics[width=\columnwidth,height=0.4\textheight]{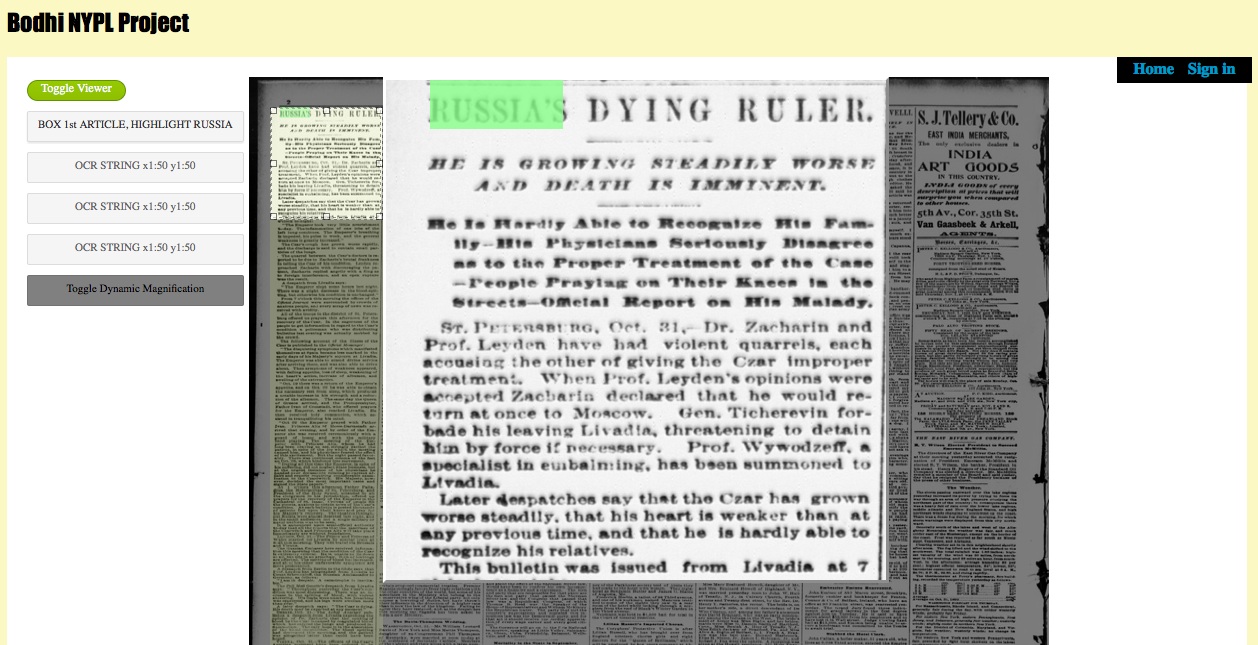}
\caption{The OCR correction module of the BODHI system, illustrating features to highlight text in articles, display high-resolution images, add tags and comments from patrons from the database.}
\label{scrShot}
\end{center}
\end{figure}

\end{document}